# High energy dissipation rates from the impingement of free paper-thin sheets of liquids: Determination of the volume of the energy dissipation zone*

Robert J. Demyanovich

*RJD Technologies, Inc., Seattle, Washington, USA*

## Abstract

The micromixing time of impinging thin liquid sheets depends upon the energy dissipation rate ($\epsilon$). The kinetic energy released by the impingement has been previously studied and was found to be a function of the coefficient of restitution of the collision. In this work, the volume within which the released kinetic energy is dissipated was investigated. The volume of energy dissipation was determined by measuring the time required for the velocity of the liquid prior to the collision to be reduced to the velocity after the collision. High-speed video was used to measure the velocity of features, generated in the front single sheet, as they passed through the impingement zone and into the mixed sheet. The experimental results showed that the time required for the velocity change was approximately equal to the residence time of liquid in the impingement zone ($t_r$). A new equation for $\epsilon$ was developed and compared with $\epsilon$ derived from turbulence energy-cascade theory. This comparison showed that the large-eddy turnover time ($t_\Lambda$) was approximately equal to $t_r$; a result that is in accordance with the notion from turbulence energy-cascade theory that large, energy-containing eddies lose their energy within $t_\Lambda$. Within the impingement zone, the large-eddy kinetic energy was found to decay exponentially with time.

## 1. Introduction

Product distributions of fast, complex chemical reactions can be significantly impacted by the rate of mixing of the reactants. If the reactants are homogenized on the molecular scale at a rate that is sufficiently fast relative to the reaction rates, the final product distribution and yield are solely dependent upon the reaction kinetics. If the reactants are segregated on the molecular scale due to poor mixing, however, the quality of chemical products can suffer. Some examples where mixing on the molecular scale (micromixing) is important include reaction injection molding (Gomes et al., 2016; Tucker and Suh, 1980), production of dyestuffs that employ fast, multiple-pathway reactions (Bourne, 2003), and molecular weight distributions of polymer molecules (Villermaux and Blavier, 1984). The rate of micromixing also impacts the particle size distribution when precipitating pharmaceuticals such as Lovastatin (Mahajan and Kirwan, 1996). With liquid rocket propellants, efficient propellant mixing and atomization improves combustion efficiency resulting in more compact combustion chambers (Halls et al., 2015).

Chemical reaction is a molecular-level process and is directly impacted by mixing on the molecular scale. “Micromixing” refers to the final mixing of reactants by molecular diffusion, which occurs within fluid elements and ultimately allows chemical reactions to proceed. Many micromixing models employ lamellar structures, where each structure initially contains only a pure reactant, in which molecular diffusion of reactants is the final step leading to chemical reaction (Baldyga and Bourne, 1984a, 1984b, 1984c; Ottino et al., 1979; Tucker and Suh, 1980). These models use a fluid mechanical approach to estimate the length scales of these structures. The models typically include thinning, stretching and folding of the structures due to fluid breakage and deformation. Since the current understanding of micromixing is based on fluid mechanics, specifically turbulence, one of the most important parameters is the energy dissipation rate ($\epsilon$) in addition to the kinematic viscosity ($\nu$).

The impingement of free thin liquid sheets (Demyanovich, 1988) is an effective method for producing high rates of energy dissipation and has been shown to yield rapid micromixing of relatively large flowrates of liquids (Demyanovich and Bourne, 1989). The liquids are formed into continuous thin sheets and impinged at one another producing a combined (or mixed) sheet of the liquids. The micromixing time for low-viscosity liquids at single-sheet flowrates of 1 to 3 L/min was found to be of order 1 to several ms. On the pilot-plant scale, yields of two commercial dyestuffs were increased by 7% and 14% relative to conventional technology (Demyanovich, 1991a, 1991b). Other potential applications include the injection of liquid rocket propellants, rapid precipitation of shear-sensitive solids and efficient carbonation of beverages (Demyanovich, 1991c).

A perspective view of the impingement of equal liquid streams is shown in Fig. 1. The more familiar case of two equal impinging jets is illustrated in Fig. 1a, while Fig. 1b depicts the impingement of two equal single sheets. For impinging jets, the collision of the jets produces a liquid sheet, whereas for impinging sheets, the collision of the single sheets produces a mixture of the liquids, which is also in the form of a continuous sheet (mixed sheet). The plane of the liquid sheet after impingement is determined by the momentum of the jets or the single sheets. When the jets or single sheets are equal, the liquid sheet or mixed liquid sheet is formed in a plane that is a bisection of the impingement angle. At equal flowrates, the thickness of the single sheets at impingement is roughly 25 times less than the diameter of the equivalent jets.

Flow within liquid sheets occurs along radial lines emanating from the origin of the single sheet with no crossover of liquid from one radial line to an adjacent radial line (except within the energy dissipation zone). Since the flowrate is constant and the velocity of sheets is constant (discussed in more detail below), liquid sheets thin with distance from the origin for single sheets and for mixed sheets with distance from the impingement zone. Ultimately, liquid sheets breakup into droplets via different mechanisms (Lin, 2003).

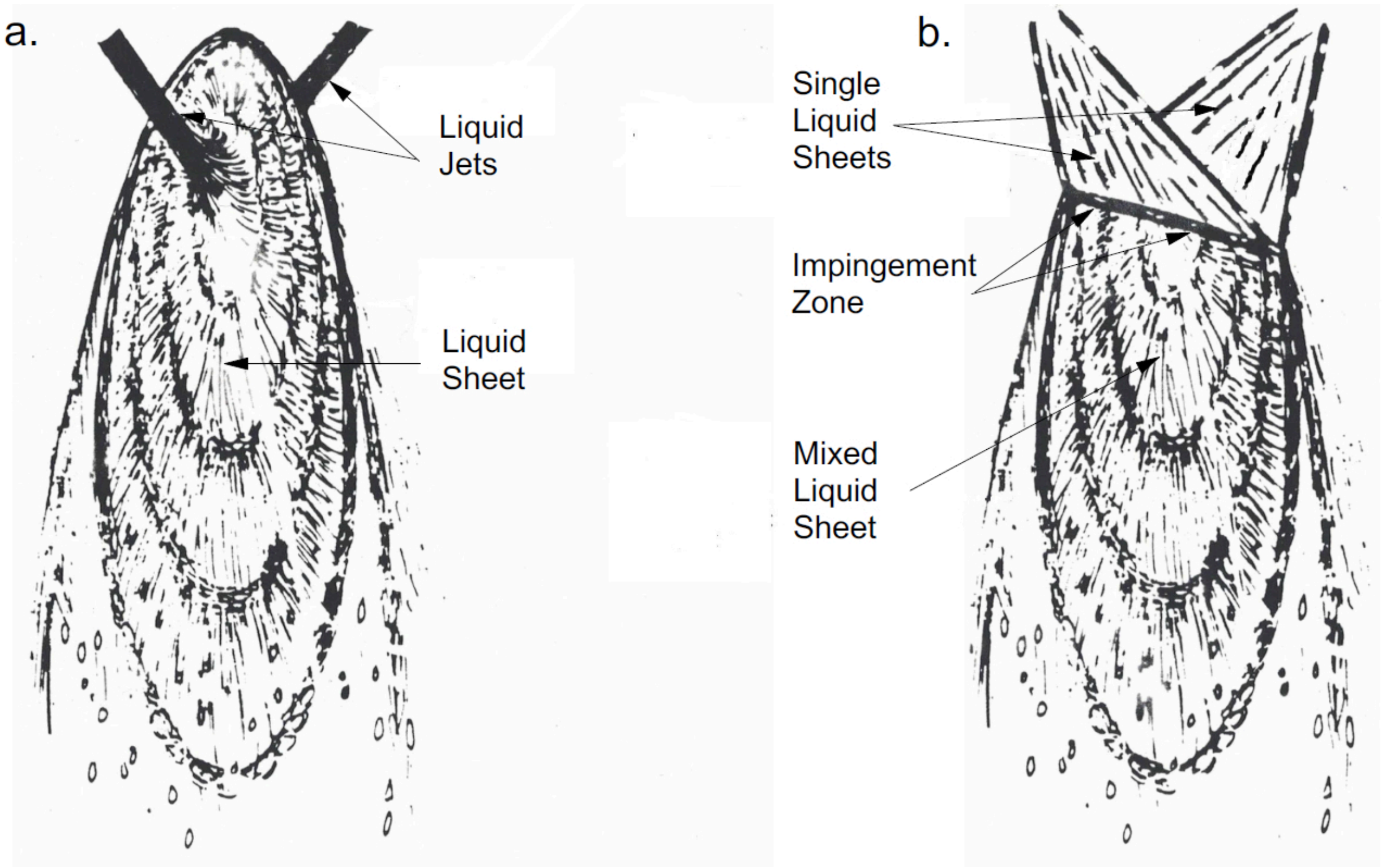

**Fig. 1.** Impingement of two equal streams of liquid. Fig. 1a is a perspective view of the impingement of two equal cylindrical jets while Fig. 1b is a perspective view of the impingement of two equal thin sheets. At impingement the thickness of the single sheets is approximately 1/25th the diameter of the equivalent jet.

As research on impinging sheets has progressed, the conceptual side view of impinging sheets has changed as shown in Fig. 2. Fig. 2a is the original side view used to determine the energy dissipation rate in the first micromixing study (Demyanovich and Bourne, 1989). It is a simplified drawing that assumes impingement does not produce a backward mixed sheet or backflow.

Recently, the amount of energy released by the collision of impinging sheets was studied leading to a revision of the side view of impingement (Demyanovich (2021a) – subsequently referred to as the $COR$ study). In Fig. 2b, the collision produces a stagnation line (shown as point SP in this cross-sectional side view), which results in deflection of flow in the forward and backward directions. In both illustrations the impingement angle is $2\beta$, the thickness of each single sheet just prior to collision is $s_i$, the distance from the origin of the single sheets to the impingement zone is $R_i$, and the velocity of each single sheet is $u$ with component velocities of $v_y$ and $v_x$. The mixed-sheet velocity is designated as $v_m$. The thickness of the forward mixed sheet is $s_f$ and, in Fig. 2b, the thickness of the backward mixed sheet is $s_B$.

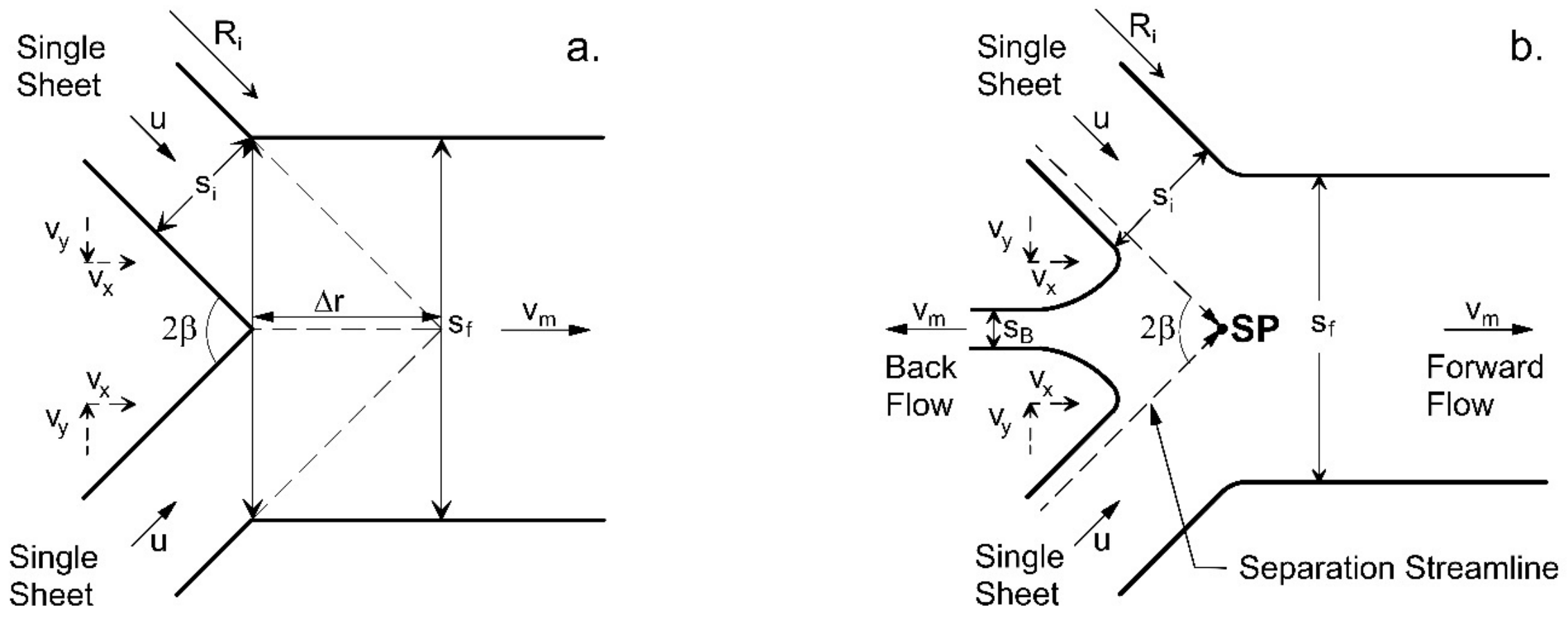

**Fig. 2.** Conceptual side view of the impingement of two equal free sheets of liquid at an angle of 2β. Fig. a is the original side view from the first micromixing study on impinging sheets, which shows a collision that does not produce a backward mixed sheet (Demyanovich and Bourne, 1989). Fig. b is an update taken from the $COR$ study, which now includes a backward mixed sheet and backflow as a result of the stagnation point SP.

For impinging sheets, the energy dissipation rate ($\epsilon$) is given by:

$$\epsilon = P/\rho V \quad (1)$$

where $P$ is the kinetic energy per unit time that is lost from the single sheets as a result of the collision, $\rho$ is the liquid density, and $V$ is the volume within which the kinetic energy is dissipated. The numerator of Eq. 1 has been previously studied in terms of the coefficient of restitution of the collision ($COR$ study). The coefficient of restitution ($COR$) for equal impinging sheets is defined as

$$COR = v_m/u \quad (2)$$

The single-sheet velocity, $u$, is known to be constant everywhere in the single sheet at ambient pressures < 20 bar (Clanet and Villermaux, 2002; Dombrowski et al., 1960; Dombrowski and Hooper, 1962; Villermaux et al., 2013). For single sheets, the velocity can be calculated as follows:

$$u = E\left(\frac{2\Delta P}{\rho}\right)^{0.5} \quad (3)$$

where $\Delta P$ is the pressure drop through the impinging-sheet device and $E$ is the efficiency of pressure head conversion to kinetic energy and ultimately velocity.

In the $COR$ study, the velocity of the mixed sheet, $v_m$, was also found to be constant. However, the experimental technique used in the $COR$ study was not able to measure mixed-sheet velocities any closer than 2.4 mm from the impingement zone (equal to approximately ten lengths of the impingement zone).

The energy per unit time released from both single sheets is readily calculated from the difference in $u$ and $v_m$ as

$$P = \frac{2\dot{m}u^2}{2} - \frac{2\dot{m}v_m^2}{2} = \dot{m}u^2 - \dot{m}v_m^2 = \dot{m}u^2(1 - COR^2) \qquad (4)$$

where $\dot{m}$ is the mass flowrate of a single sheet and $2\dot{m}$ is the mass flowrate of the mixed sheet. The velocity ($v'$) associated with the kinetic energy released as a result of the collision is given by:

$$v' = u\sqrt{1 - COR^2} \qquad (5)$$

With respect to the release of kinetic energy, the impingement of liquid sheets results in two types of collisions: elastic collisions and inelastic collisions ($COR$ study, Demyanovich (2021b)). A theoretically, perfectly elastic collision is one in which kinetic energy is not released from the single sheets and the mixed-sheet velocity is equal to the single-sheet velocity. An inelastic collision of equal impinging sheets is a collision that results in the dissipation of the kinetic energy from the single-sheet component velocities ($v_y$ in Fig. 2) that are destroyed upon impact. It does not refer to a collision where all the kinetic energy of the single sheets is dissipated upon impingement; it does not seem possible to impinge two liquid sheets in this manner. In this case, the mixed-sheet velocity ($v_m$) is theoretically equal to the x-component velocity of the single sheets ($v_x = ucos\beta$). For an elastic collision the $COR$ is between the $COR$ for an inelastic collision ($COR = cos\beta$) and a perfectly elastic collision ($COR$ = 1). With elastic collisions some of the kinetic energy associated with $v_y$ is restored after the collision.

In the original micromixing study on impinging sheets (Demyanovich and Bourne, 1989), the volume of the energy dissipation zone in Eq. 1 was assumed to be equal to the impingement zone volume. The volume was determined from the geometry of Fig. 2a and resulted in calculated energy dissipation rates between $2 \times 10^3$ and $4 \times 10^5$ W/kg. A recent article on the micromixing of free impinging jets (see Fig. 1a) used two methods to calculate the energy dissipation rate: one based on the jet impingement zone volume and the other based on the entire volume of the formed liquid sheet (Abiev and Sirotkin, 2020). These authors were uncertain which volume should be used for calculating the energy dissipation rate of impinging jets. The difference in the energy dissipation rates calculated using these different volumes is a factor of $10^2$ to $10^4$.

The $COR$ study established the amount of energy released as a result of the impingement (the numerator in Eq. 1) but was not able to provide an estimate of the mass within which that energy is dissipated. Calculation of this mass has, up to now, assumed that the energy is dissipated within a volume equal to the impingement zone volume. In light of this assumption and the uncertainty raised in the recent impinging-jet micromixing study (Abiev and Sirotkin, 2020), the focus of this work is to more precisely determine the volume of the energy dissipation zone and to update the expression for the energy dissipation rate of impinging sheets.

The remainder of this article is structured as follows: Section 2 discusses the experimental method used to measure displacements and velocities in the energy dissipation zone. Results and discussion are presented in Section 3. Based on the experimental results, a new derivation of the volume of the impingement zone is presented in Section 3.2 in terms of readily obtained or calculated parameters. Section 3.2.1 compares the new method for calculating the energy dissipation rate ($\epsilon$) with the previous method. To support the analysis of micromixing studies using a fluid mechanical approach, Section 3.3 compares $\epsilon$ derived from Eq. 1 with $\epsilon$ derived from turbulence energy cascade theory. Such a comparison can assist with the calculation of turbulence parameters, such as large-eddy size, large-eddy turnover time, turbulent kinetic energy, turbulence intensity, large-scale eddy Reynolds number, etc. Section 3.4 looks at the kinetic energy profile in the energy dissipation zone. Section 4 provides a short summary and conclusions.

## 2. Experimental

The energy dissipation zone is the mass within which the kinetic energy released by the collision is dissipated. Since the velocity in the single and mixed sheet is constant, liquid at the beginning of the energy dissipation zone has a velocity equal to $u$ and at the end of the energy dissipation zone a velocity equal to $v_m$.

Velocities in the impingement zone and mixed sheet were experimentally measured from high-speed video of the displacement of features, which form in a liquid sheet when a soluble or emulsified oil is added to the liquid. In the $COR$ study, sheet velocities were measured from double-exposure photographs of the movement of holes over a fixed time interval (see Fig. 3). However, holes were seldom observed to form prior to impingement. If velocities within the impingement zone are to be measured, then features must be generated before the collision in order to view these features as they pass through the impingement zone. It was found that commercially available bleach, containing emulsified lavender oil, consistently generated features prior to impingement.

Sheet velocities of features created in 48.5% w/w (6 mPa s) and 65.8% w/w (26 mPa s) aqueous glycerine were measured. Higher viscosities dampen impact waves yielding features in the impingement zone and mixed sheet with less distortion, and thereby increasing the accuracy of measurements. Viscosities up to 70 mPa s have been found to have no effect on the constancy in velocity within a liquid sheet (Dombrowski et al., 1960). Further, the $COR$ study demonstrated that the $COR$ is not a function of surface tension for impinging sheets of the same surface tension. Standard values of the liquid physical properties were taken from the literature. The temperature for all experiments was 20$^{o}$C $\pm$ 2$^{o}$C. Methods for creating and impinging sheets are provided elsewhere (Demyanovich, 1988).

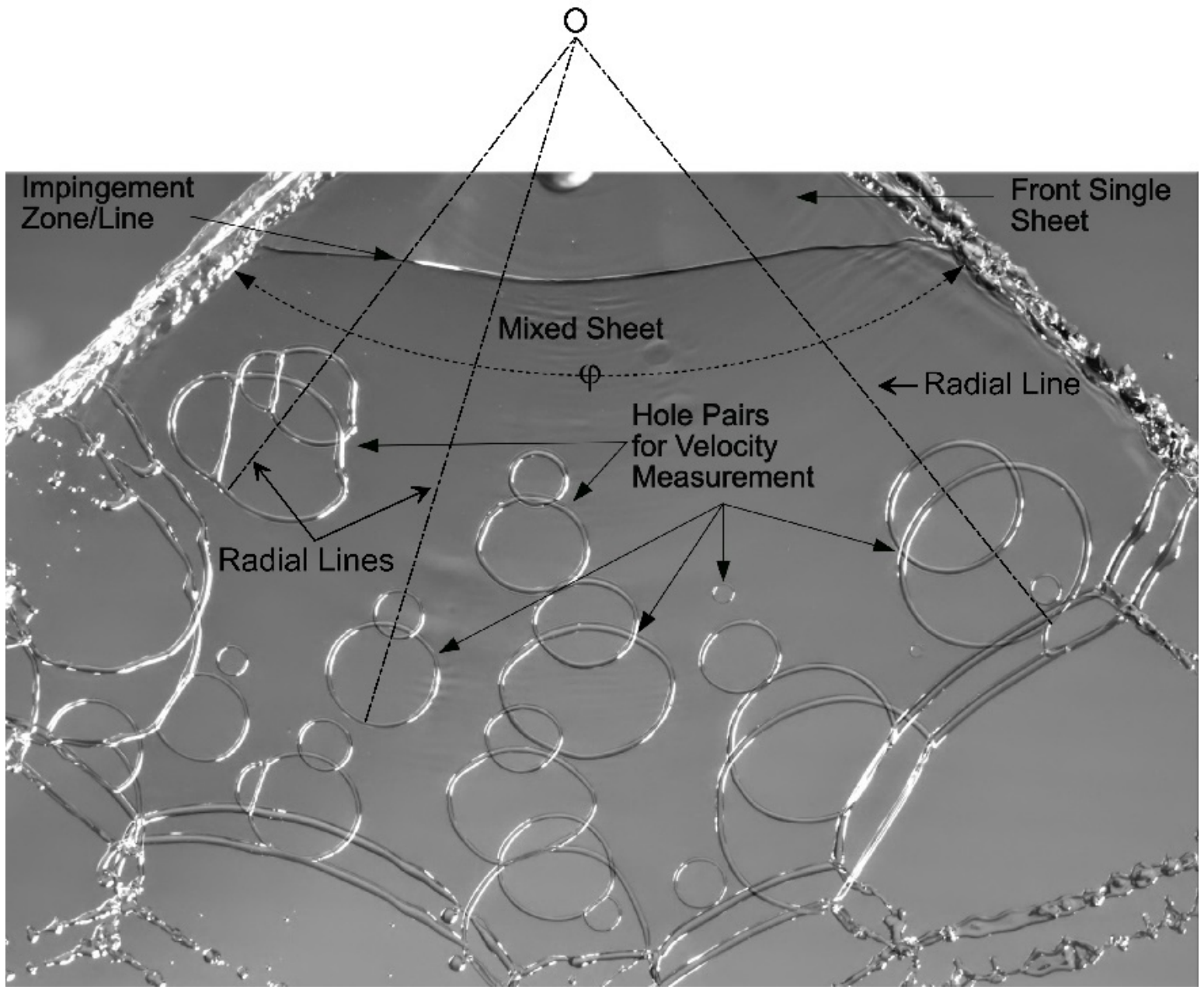

**Fig. 3.** Double-exposure photograph of the inelastic collision of two equal single sheets of 32% aqueous glycerin at an impingement angle of 44° and pressure drop of 0.195 bar (single-sheet velocity calculated from Eq. 3 is 4.76 m/s). The duration of each flash was 4 μs with a delay time between flashes of 1.0 ms. Mixed-sheet velocities were measured for six hole pairs. The mean mixed-sheet velocity was 4.33 m/s with a standard deviation of 2.2%. The mean coefficient of restitution (*COR*) was calculated from Eq. 2 as 0.910 ($\cos\beta$ = 0.927). $\varphi$ is the spread angle of the sheets. Radial lines for three of the six hole pairs are shown on the photograph. These radial lines converge at the origin of the fan-shaped, single sheets ("O"). Despite the expansion of the holes over time, the center of each hole pair remains on the radial line during its lifespan within the mixed sheet. This figure is a modification of the Graphical Abstract in the $COR$ study.

The experimental setup is illustrated in Fig. 4. It is the same apparatus as that used in the $COR$ study except that the camera is now a video camera instead of a still camera and the light source is a 300 W LED light instead of a microsecond flash unit. Liquids were pumped from storage tanks (ST) via gear pumps (P) driven by variable speed drives (VSD). The positions of input selector valves (ISV) were adjusted to either direct liquid from the storage tanks to the impinging sheet device or to flow tap water through the apparatus for video camera setup and system cleaning. Liquid flow rates were measured by rotameters (R) and the pressure drop through the impinging-sheet device was measured using digital pressure gauges (PG) accurate to 1% of full scale. After liquid flowed to the impinging sheet device, droplets from the mixed sheet (MS) were collected in a tank (CT).

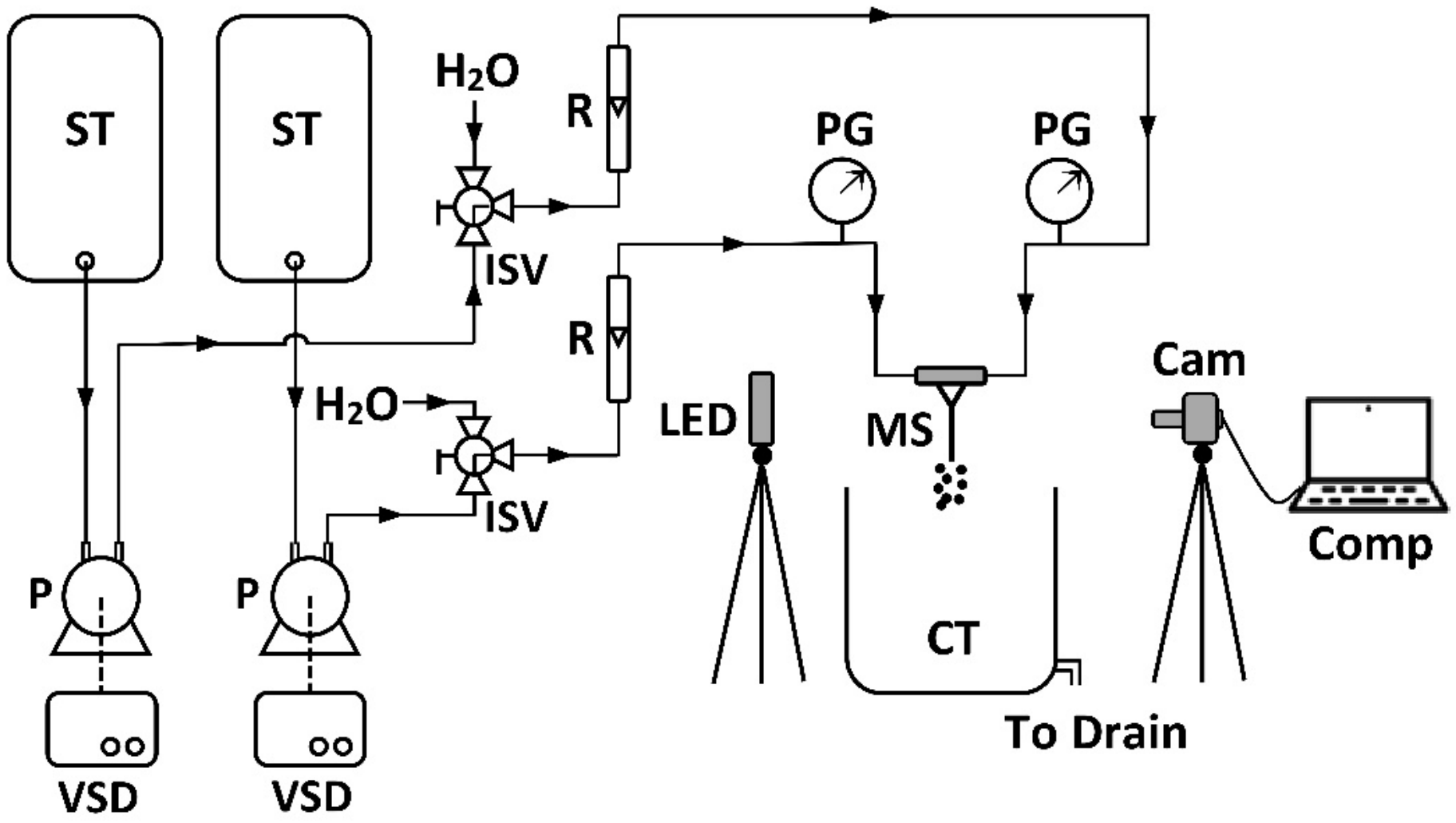

**Fig. 4.** Illustration of the experimental apparatus. ST – storage tank, P – gear pump, VSD – variable speed drive, ISV – input selector valve, R – rotameter, PG – pressure gauge, MS – mixed sheet, CT – collection tank, LED – 300W LED light source, Cam – video camera, Comp – laptop computer.

Two to three-second videos were taken at speeds of 24,047 frames per second. Short clips of a few milliseconds in duration of the movement of a feature from the single sheet, through the impingement zone, and into the mixed sheet were extracted for post processing and subsequent velocity measurement. Using video-editing software, post processing typically involved increasing exposure, converting to a color scheme that highlighted the centers of the features when applicable, and sharpening the video the maximum amount possible.

## 3. Results and discussion

### *3.1 Measurement of liquid displacement and velocity in the impingement zone*

Experimental features were generally of two types: expanding holes and non-expanding holes. Fig. 5 contains extracted frames from a video sequence of the movement of an expanding hole through the impingement zone, while Fig. 6 contains extracted frames from a video sequence of a non-expanding hole.

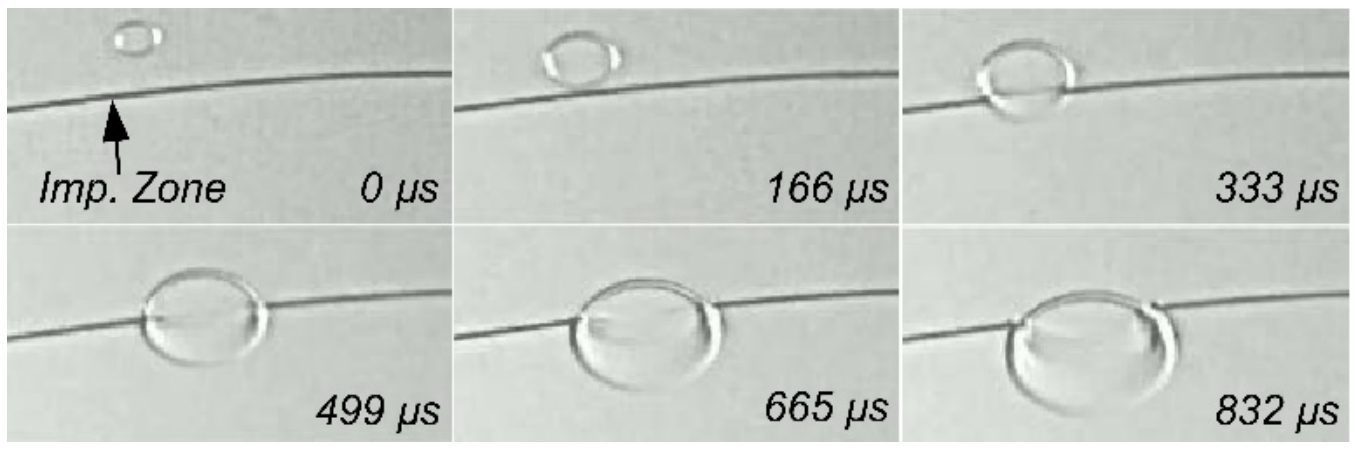

**Fig. 5.** Movement of an expanding hole formed in the front single sheet. After ~0.5 ms, the hole center has passed through the impingement zone and into the mixed sheet. Imp. Zone – impingement zone.

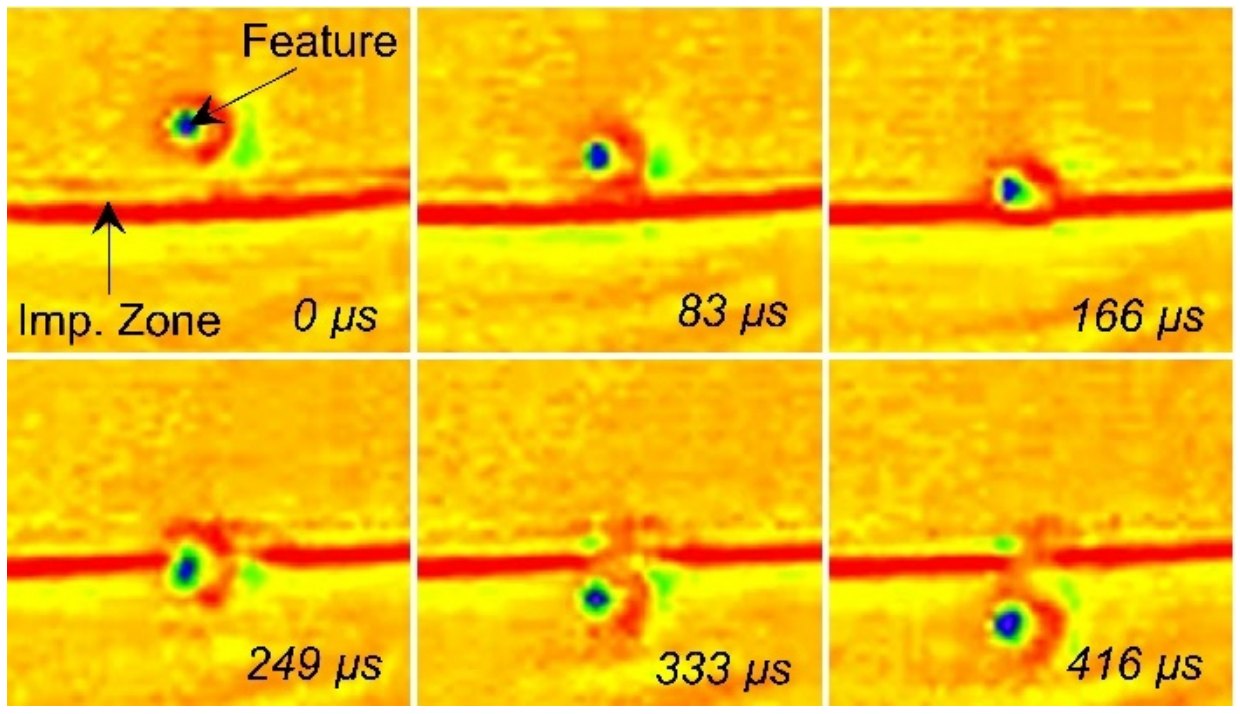

**Fig. 6.** Movement of a non-expanding hole or feature, formed in the front single sheet, through the impingement zone and into the mixed sheet. Image has been color processed to highlight the center of the feature.

ImageJ software (Schneider et al., 2012) was used to determine the center of the feature in each frame. For the non-expanding hole shown in Fig. 6, Fig. 7 illustrates the path that the center (the darkest portion of the non-expanding hole) took during about 0.4 ms of movement. The movement of the center traced the path of the feature as it passes from the single sheet through the impingement zone ($IZ$) and into the mixed sheet. ImageJ provided the coordinates of each movement of the hole center with time, which were exported to a spreadsheet. These coordinates were measured in pixels by ImageJ, and frame-to-frame pixel displacement was converted to mm by a reference measurement from a ruler (with 0.5 mm divisions) on each video clip. $\Delta b$ is the new designation for the length of the impingement zone and is generally not equal to $\Delta r$ shown in Fig. 2a.

Results such as Fig. 7 indicate that the feature center appeared to remain approximately on the same radial line throughout the movement. However, since energy is known to be released and likely results in some degree of turbulence, it is unlikely that the center remained on a radial line in the energy dissipation zone. Rather, the sensitivity of this experimental technique is insufficient to measure very small movements (probably on the order of 50 μm or less) in the azimuthal direction ($\varphi$ in Fig. 3).

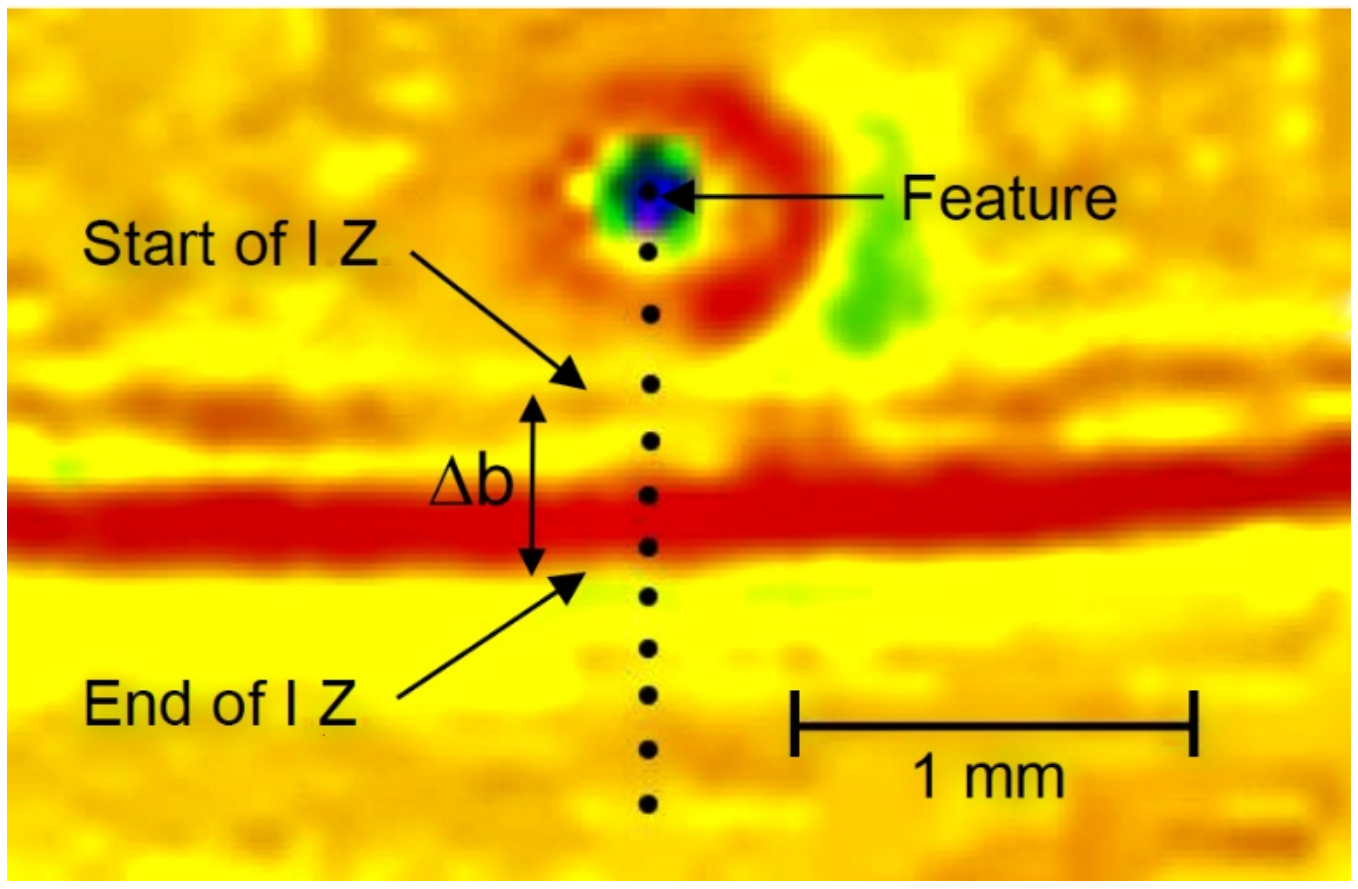

**Fig. 7.** Sequence of movement of the hole shown in Fig. 6. The movement of the hole center is shown as black dots. $\Delta b$ – length of the impingement zone.

From a two to three-second video, multiple different feature sequences were extracted for measuring movement. Within each feature sequence between 11 to 25 displacements of the feature were measured (only six are shown in Fig. 6 whereas all 11 are shown in Fig. 7). In Fig. 7, the reddish horizontal band labelled $IZ$ (impingement zone) is the projection of the front single sheet onto the mixed sheet. Measurement of hole displacement was separated as movement in the single sheet, movement in the impingement zone and movement in the mixed sheet. Only the impingement zone and mixed sheet were perpendicular to the video camera.

The experimental parameters are listed in Table 1. For each run the mean $v_m$ is an average of the velocities measured in the mixed sheet (not including the impingement zone) for the number of feature sequences shown in column 5. Since the experimental viscosities had not been previously studied, values of $E$ needed for the calculation of the single-sheet velocity from Eq. 3 were not available. Therefore, the values of $u$ listed in Table 1 were directly measured from the movement of holes in the single sheets. The experimental $COR$ was determined from Eq. 2.

For all runs the collisions of the impinging sheets were inelastic. Elastic collisions were not investigated because the backflow of liquid obscures perforations when they are close to or in the impingement zone. For inelastic collisions the $COR$ is theoretically equal to $cos\beta$. A comparison of columns 11 and 12 shows good agreement between the measured $COR$ and the theoretical $COR$. As found in the $COR$ study, however, the experimental $COR$ was somewhat lower for inelastic collisions at higher viscosity/density when compared with water.

Column 6 in Table 1 lists the number of displacement vectors for which a length could be measured within the impingement zone. This number is a function of the single-sheet thickness at impingement ($s_i$), half impingement angle ($\beta$), velocity in the impingement zone, and the camera frames per second. The range of single-sheet thicknesses at impingement ($s_i$) was 140 µm to 347 µm. The standard deviation of length measurements within the impingement zone ranged from 2.7% to 12.6% of the mean value (mean for the number of feature sequences listed in column 5).

**Table 1.** Parameters for the displacement measurement experiments. $2\beta$ – impingement angle; $COR$ – coefficient of restitution; $\Delta b$ – impingement zone length; exp. – experimental; feat. – feature; IZ – impingement zone; meas. – measured; Nom. – nominal; $\nu$ – viscosity; Nbr. – number; ΔP – pressure drop; seq. – sequences; % st. dev. – standard deviation divided by the mean and reported as a percentage; $u$ – single-sheet velocity; $v_m$ – mixed-sheet velocity.

| Run | 2β (°) | Nom. ΔP (kPa) | $\nu$ x10$^{-6}$ (m$^2$/s) | Nbr. of feat. seq. | Nbr. of vectors meas. in IZ per seq. | Mean $u$ (m/s) | % st. dev. for mean $u$ | Mean $v_m$ (m/s) | % st. dev. for mean $v_m$ | Mean exp. $COR$ | cos$\beta$ | Mean $\Delta b$ (μm) | % st. dev. for mean $\Delta b$ |
|---|---|---|---|---|---|---|---|---|---|---|---|---|---|
| 1 | 46 | 12.0 | 5.3 | 15 | 3 | 3.56 | 0.9% | 3.23 | 6.3% | 0.91 | 0.92 | 312 | 10.0% |
| 2 | 46 | 19.5 | 5.3 | 15 | 2 | 4.57 | 1.2% | 4.09 | 3.6% | 0.89 | 0.92 | 361 | 7.4% |
| 3 | 75 | 15.0 | 22.3 | 10 | 3 | 3.37 | 1.4% | 2.47 | 8.4% | 0.73 | 0.79 | 333 | 6.6% |
| 4 | 75 | 10.0 | 22.3 | 5 | 4 | 2.75 | 2.5% | 2.02 | 12.0% | 0.74 | 0.79 | 379 | 9.0% |
| 5 | 67 | 10.7 | 5.3 | 10 | 3 | 3.08 | 2.3% | 2.63 | 7.2% | 0.86 | 0.83 | 367 | 10.1% |
| 6 | 67 | 12.0 | 5.3 | 8 | 3 | 3.26 | 0.7% | 2.65 | 6.2% | 0.81 | 0.83 | 368 | 6.8% |
| 7 | 67 | 17.8 | 5.3 | 8 | 2 | 3.95 | 1.1% | 3.12 | 5.5% | 0.79 | 0.83 | 255 | 6.1% |

Measurements of the length ($\Delta b$) of the impingement zone were made for each sequence within an experimental run. For a specific run the radial lines containing the features were, in general, not identical (features were located at random locations in the single sheet similar to Fig. 3). Consequently, the impingement zone length varied for the sequences within a run. The standard deviation in the measurement of $\Delta b$ for all runs ranged between 6.1% to 10.1%.

With a focus on the impingement zone, the measured displacements ($S$) of the features were plotted on position versus time ($t$) charts. The experimental results for each run were fit to 84 different regression models. Of these regression models, three were found to have regression coefficients ($R^2$) that were consistently ≥ 0.9997 for all experimental runs. The three regression models are hyperbolic decline, exponential association and rational model. While there were a few additional regression models that consistently provided excellent fits, these three were chosen because they represent a decline model, a model based on an exponential and a model that requires 4 parameters for a fit. Further, the derivatives of these model equations can be determined at $t = 0$, which was not the case for some of the other, high-$R^2$ regression models. The equations are

hyperbolic decline:

$$S = f\left(1 + \frac{ht}{g}\right)^{-\frac{1}{h}} \tag{6}$$

exponential association:

$$S = f(g - \exp(-ht)) \tag{7}$$

rational model:

$$S = \frac{f + gt}{1 + ht + jt^2} \tag{8}$$

where f, g, h and j are coefficients specific to each regression.

In Fig. 8 the experimental results along with the fits for the three regression models are plotted for Run 6. The plot shown in Fig. 8 is typical of all runs. The fits of the three equations to the data are excellent and indistinguishable. The velocity at any specific time is equal to the slope of the curve. Where the slope equals the mixed-sheet velocity is the time ($t_c$) at which the sheet velocity has been reduced from $u$ to $v_m$ as a result of the loss of kinetic energy. $\Delta w$ is the distance liquid has travelled during the time $t_c$. Whereas $\Delta b$ represents the length of the impingement zone, $\Delta w$ represents the length of the energy dissipation zone. $\Delta w$ could be less than, equal to, or greater than $\Delta b$. The time at which $S$ - $S_i$ is equal to $\Delta b$, where $S_i$ is the value at $t = 0$, is the residence time ($t_r$) of liquid in the impingement zone.

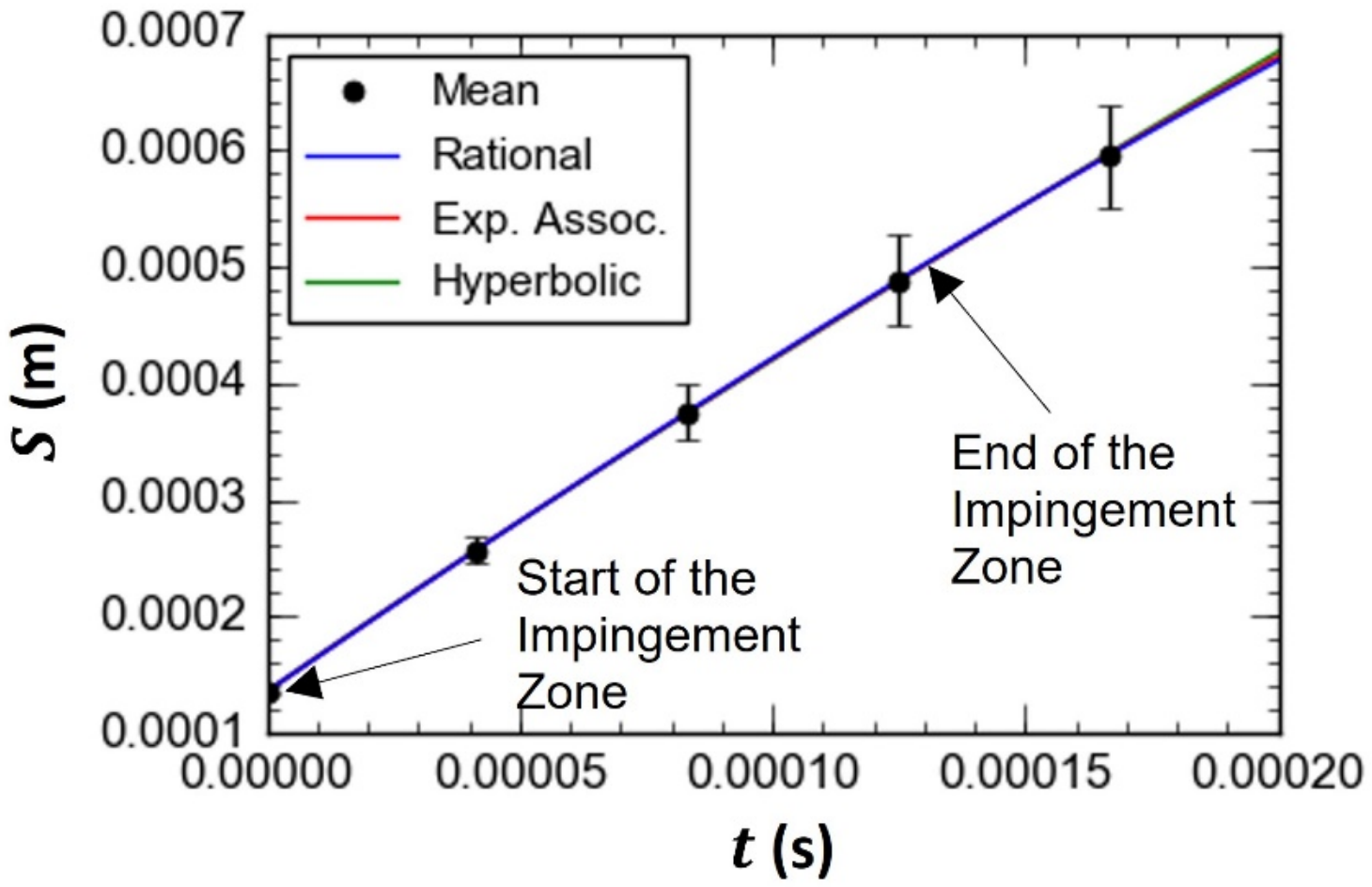

**Fig. 8.** Plot of displacement ($S$) versus time ($t$) for Run 6. The fits for the three regression equations are shown and are indistinguishable from one another. Error bars represent the standard deviation in measurements of the displacements relative to the plotted mean values. For Run 6 the standard deviations illustrated by the error bars ranged from 4.0% to 7.5% of the plotted mean value.

For determining $t_c$, the instantaneous flow velocity in the impingement zone ($\hat{u}$) was calculated from the derivative of the equations for the regression fits:

hyperbolic decline:

$$\hat{u} = \frac{dS}{dt} = -fg^{1/h}(g + ht)^{(-1-h)/h} \tag{9}$$

exponential association:

$$\hat{u} = \frac{dS}{dt} = fh\exp(-ht) \tag{10}$$

rational model:

$$\hat{u} = \frac{dS}{dt} = \frac{g - fh - 2fjt - gjt^2}{(1 + ht + jt^2)^2} \quad (11)$$

As shown in Fig. 8, the experimental data was subject to error. When a differential operator is applied to discrete position versus time data, errors tend to be magnified (Brown et al., 1992). Therefore, the calculation of $t_c$ was performed for the mean values of the position data (shown as "Mean" in Fig. 8) and not for values from each individual sequence (for each of the eight sequences of Run 6 for example). Applying the differential operators to the mean helped to supress the magnification of the experimental errors.

Once a value of $t_c$ is calculated for a regression model, the corresponding displacement, $S_c$, can be determined from the displacement vs time equation for the particular regression model (Eq. 6, Eq. 7, or Eq. 8). $\Delta w$ is then equal to $S_c$ - $S_i$, where $S_i$ is the value at $t = 0$.

For all runs, the calculated values of $t_c$ and $\Delta w$ determined from each regression model are listed in Table 2. For all regression equations, the ratio, $\Delta w/\Delta b$, is sufficiently close to 1.0 to conclude that the distance or length ($\Delta w$) required for the velocity to change from the single-sheet velocity ($u$) to the mixed-sheet velocity ($v_m$) is equal to the length of the impingement zone ($\Delta b$). As will be shown in the next section, this implies that the volume of the energy dissipation zone is equal to the volume of the impingement zone.

**Table 2.** Calculations of the length of the energy dissipation zone for each run in Table 1. $\Delta b$ – impingement zone length; % S.D. – standard deviation divided by the mean and reported as a percentage; $t_c$ – time for the velocity in the energy dissipation zone to change from $u$ to $v_m$; $t_r$ – residence time of liquid in the impingement zone; $\Delta w$ – length of the energy dissipation zone or length at which the velocity has decreased from $u$ to $v_m$.

| | | | **Hyperbolic decline** | | | **Exponential Association** | | | **Rational model** | | |
|---|---|---|---|---|---|---|---|---|---|---|---|
| **Run** | **$t_r$ (μs)** | **$\Delta b$ (μm)** | **$t_c$ (μs)** | **$\Delta w$ (μm)** | **$\Delta w/\Delta b$** | **$t_c$ (μs)** | **$\Delta w$ (μm)** | **$\Delta w/\Delta b$** | **$t_c$ (μs)** | **$\Delta w$ (μm)** | **$\Delta w/\Delta b$** |
| 1 | 94 | 312 | 74 | 251 | 0.80 | 83 | 280 | 0.90 | 65 | 222 | 0.71 |
| 2 | 84 | 361 | 86 | 369 | 1.02 | 88 | 380 | 1.05 | 85 | 367 | 1.02 |
| 3 | 118 | 333 | 114 | 325 | 0.97 | 116 | 331 | 0.99 | 105 | 301 | 0.90 |
| 4 | 167 | 379 | 158 | 362 | 0.95 | 153 | 355 | 0.94 | 154 | 353 | 0.93 |
| 5 | 127 | 367 | 118 | 343 | 0.93 | 121 | 351 | 0.96 | 93 | 275 | 0.75 |
| 6 | 131 | 368 | 135 | 379 | 1.03 | 128 | 361 | 0.98 | 128 | 362 | 0.98 |
| 7 | 75 | 255 | 79 | 267 | 1.05 | 83 | 279 | 1.10 | 74 | 250 | 0.98 |
| Mean | | | | | **0.97** | | | **0.99** | | | **0.90** |
| % S.D. | | | | | 8.6% | | | 6.9% | | | 13.4% |

The plot of the calculated velocity profile for Run 6 is shown in Fig. 9. For the interior points (which don't include $t = 0$ and $t = 0.166$ ms), all three models more or less calculated the same value (within ~ 1.4% for Run 6 and ≤ 4.7% for all runs). At the end points (single-sheet velocity at $t = 0$ and mixed-sheet velocity at the greatest time), no particular model provided better estimates of the experimental values (▲) than the other models. However, the regression models typically did not agree as well with one another when calculating the single and mixed-sheet velocities. The range of differences when calculating the single-sheet velocity was from 2.5% to 12.5% with a mean of 6.9%. On average, this mean was about 3.9% different than the experimentally measured single-sheet velocity. The range of differences when calculating the mixed-sheet velocity was from 1.5% to 12.0% with a mean of 6.5%. On average, this mean was about 4.4% lower than the experimentally measured mixed-sheet velocity.

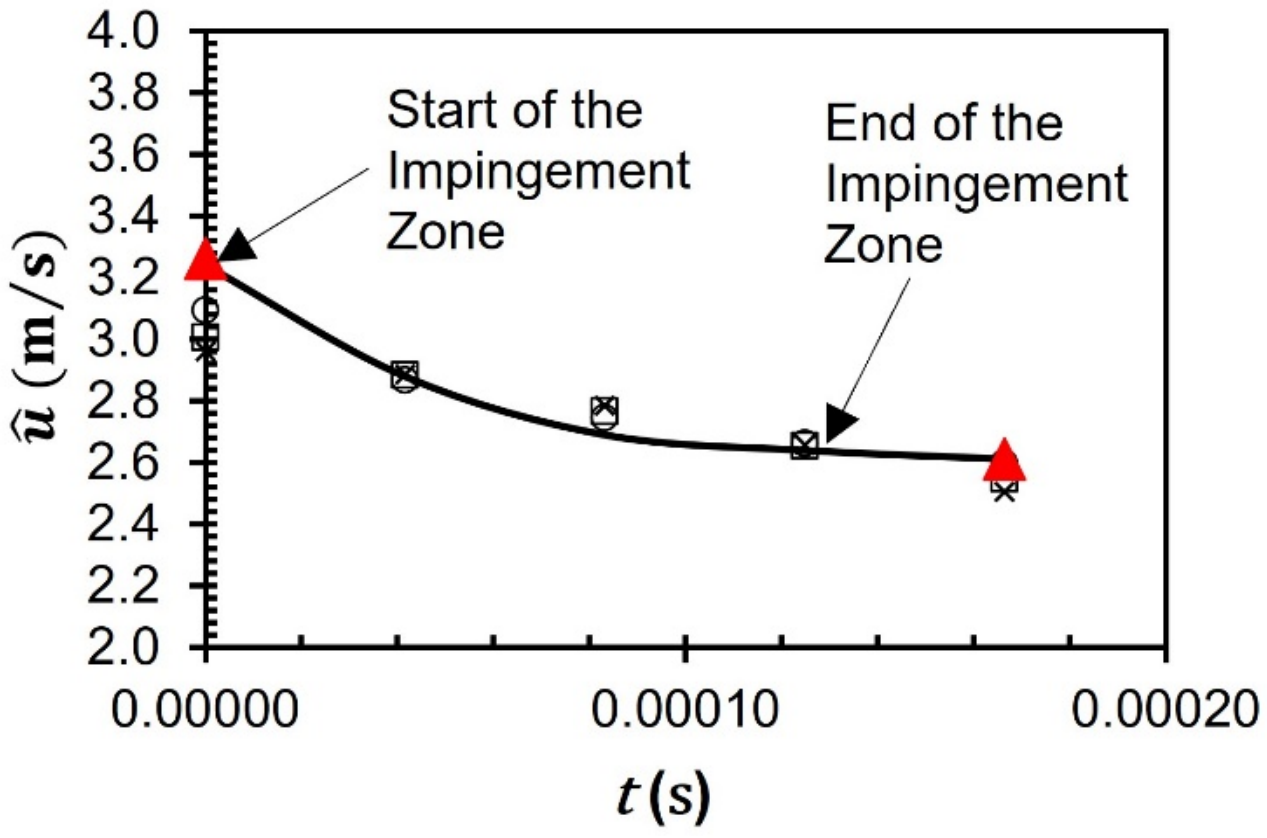

**Fig. 9.** Plot of flow velocity in the impingement zone ($\hat{u}$) versus time for Run 6. Velocities were calculated from the derivatives of the three regression equations. ▲ experimentally measured single-sheet and mixed-sheet velocities; **O** hyperbolic decline; **✕** rational model; □ exponential association. Line intended as a visual aid only.

## *3.2 Impingement zone volume and energy dissipation rate*

In Table 2, the length ($\Delta w$) required for the velocity to change from $u$ to $v_m$ was equal to the length of the impingement zone ($\Delta b$). Therefore, the energy is dissipated within the impingement zone.

The volume of the impingement zone can be calculated based on the $COR$ and geometric/trigonometric considerations. As shown in Fig. 10a, the impingement zone of equal impinging sheets is a concentric shell with a volume equal to $V_{IZ}$. Since $\Delta b \ll R_i$, $V_{IZ}$ is equal to,

$$V_{IZ} = \varphi R_i A_{IZ} \tag{12}$$

where $\varphi$ is the azimuthal or spread angle of the sheets, $A_{IZ}$ is the cross-sectional area of the impingement zone, and $R_i$ is the distance from the origin of the single sheets to the impingement zone. $A_{IZ}$ is illustrated in the new diagram of the side view of impinging sheets shown in Fig. 10b.

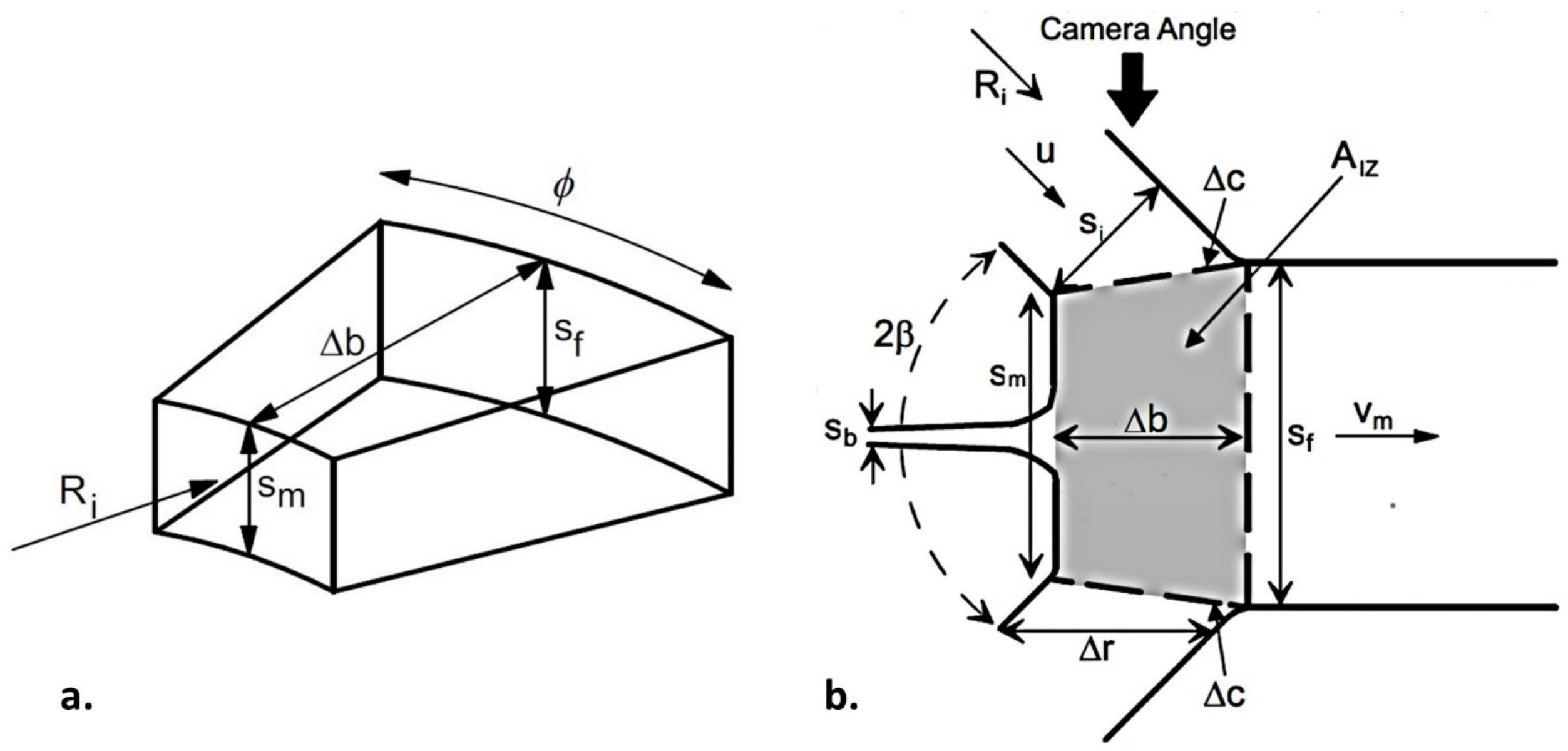

**Fig. 10.** Fig. a illustrates the impingement zone as a cylindrical shell. Fig. b is a cross section of the single sheet, impingement zone (cylindrical shell) and mixed sheet depicting a conceptual side view of the collision of equal impinging sheets. The dashed lines bound the cross-sectional area of the impingement zone which is shaded in gray. $A_{IZ}$ – cross-sectional area of the impingement zone; $\beta$ – half angle of impingement; $\Delta c$ – projection of the single sheet onto the mixed sheet; $\varphi$ – azimuthal direction; $\Delta b$ – length of the impingement zone; $\Delta r - s_i/sin\beta$; $R_i$ – radial distance from the origin of the single sheets to the impingement zone; $s_b$ – thickness at the rim of the backward mixed sheet; $s_f$ – thickness of the forward mixed sheet at the end of the energy dissipation zone; $s_i$ – single-sheet thickness at the impingement zone; $s_m$ – thickness of the mixed sheet at the beginning of the impingement zone; $u$ – single-sheet velocity; $v_m$ – mixed-sheet velocity. Figures are not to scale with each other.

Compared with the side view schematic shown in Fig. 2a, the updated diagram depicts the mixed-sheet thickness increasing within the impingement zone (from $s_m$ to $s_f$). Since the single-sheet flowrates are constant, this increase in thickness is a result of the velocity decrease. Although sheets thin as they expand, the thinning that occurs within the impingement zone is only about 1% to 2% of the thickness, $s_f$, and will be neglected.

For each experimental run, the camera was perpendicular to the impingement zone and mixed sheet as shown in Fig. 10b. Thus, the measured distance of the projection of the single sheet onto the mixed sheet ($\Delta c$) in a video was $\Delta b$. The projection of the single sheet onto the mixed sheet is the main visible clue of impingement and is referred to as the impingement zone. Since the kinetic energy per unit time in Eq. 4 was released by the time liquid had travelled the length, $\Delta b$, the impingement zone volume will now be defined as the mass of liquid between the projections, $\Delta c$, of each single sheet onto the mixed sheet.

The cross-sectional area of the impingement zone, $A_{IZ}$, is the area of a trapezoid which is given by

$$A_{IZ} = \frac{(s_m + s_f)}{2} \Delta b \quad (13)$$

The cross-sectional area ($A_{EDZ}$) of the energy dissipation zone ($EDZ$) is

$$A_{EDZ} = \frac{(s_m + s_f)}{2} \Delta w \quad (14)$$

However, since $\Delta w = \Delta b$, $A_{EDZ} = A_{IZ}$ and from Eq. 12 the volume of the energy dissipation zone is equal to the volume of the impingement zone.

In terms of the single-sheet thickness at impingement ($s_i$) and the projection ($\Delta r$) of $s_i$ at the half angle of impingement ($\beta$), Eq. 13 becomes

$$A_{IZ} = Z s_i \Delta r \quad (15)$$

Details on the derivation of Eq. 15 are given in Appendix A. Since the velocity and flowrate within liquid sheets are constant, $s_i$ is readily calculated from the flow equation for a single sheet, which is

$$\dot{m} = \rho \varphi R_i s_i u \quad (16)$$

The projection of the single sheet ($\Delta r$) into the mixed sheet (see Fig. 2a) is equal to $s_i / sin\beta$. The derivation of the geometric factor, $Z$, can be found in Appendix A. $Z$ is only a function of $cos\beta$ and $COR$. Over the range $30^o \leq 2\beta \leq 120^o$, $Z$ varies from 1.934 to 1.475 for an elastic collision and 1.916 to 1.031 for an inelastic collision. In the $COR$ study, impingement angles up to $120^o$ were investigated, which will be the assumed upper limit for the validity of the experimental results and analyses of this investigation.

Since the volume of the energy dissipation zone is equal to the volume of the impingement zone, $\epsilon$ can be derived by Substituting Eq. 4, Eq. 12, Eq. 15, Eq. 16, and Eq. 5 into Eq. 1, which yields the following expression:

$$\epsilon = \frac{P}{\rho V} = \frac{\rho \varphi R_i s_i u^3 (1 - COR^2)}{\rho \varphi R_i s_i Z \Delta r} = \frac{u^3 (1 - COR^2)}{Z \Delta r} = \frac{v'^2 u}{Z \Delta r} \quad (17)$$

For the experimental parameters listed in Table 1, the calculated values of $\epsilon$ range from $1.03 \times 10^4$ W/kg to $4.09 \times 10^4$ W/kg with a mean uncertainty of 33% (see Appendix B for the calculation of the propagation of uncertainty for Eq. 17)

### *3.2.1 Comparison with previous method for calculating energy dissipation rates*

As noted for Fig. 2a, the impingement zone was previously viewed as the projection of the single sheets ($\Delta r = s_i / sin\beta$) into the mixed sheet (and not onto the mixed sheet) with contact at the centerline of the mixed sheet. This view was based on streamline flow within the impingement zone even though it was known that energy is released as a result of the collision. Further, Fig. 2a is limited to inelastic collisions and assumes an instantaneous reduction in velocity from $u$ to $v_m$. These assumptions, however, were believed to result in conservatively high estimates of the volume of the impingement zone and, therefore, conservatively low estimates of $\epsilon$.

In the original method shown in Fig. 2a, the energy dissipation rate, $\epsilon_o$, for inelastic collisions was derived as (Demyanovich and Bourne, 1989)

$$\epsilon_o = \frac{v'^3 cos\beta}{2s_i} \tag{18}$$

For an inelastic collision $v' = usin\beta$ and since $\Delta r = s_i/sin\beta$, Eq. 18 can be rewritten as

$$\epsilon_o = \frac{v'^2 usin\beta cos\beta}{2\Delta rsin\beta} = \frac{v'^2 ucos\beta}{2\Delta r} \tag{19}$$

Thus, the ratio of Eq. 17 to Eq. 19 is

$$\frac{\epsilon}{\epsilon_o} = \frac{2}{Zcos\beta} \tag{20}$$

The ratios of the energy dissipation rates for inelastic collisions as a function of impingement angle are listed in Table 3.

**Table 3.** Comparison of the energy dissipation rate for an inelastic collision calculated from Eq. 17 with the earlier method for calculating the energy dissipation rate for an inelastic collision given by Eq. 19.

| $2\beta$ (°) | $cos\beta$ | $Z$ | $\epsilon/\epsilon_o$ |
|---|---|---|---|
| 30 | 0.966 | 1.916 | 1.08 |
| 45 | 0.924 | 1.817 | 1.19 |
| 60 | 0.866 | 1.687 | 1.37 |
| 75 | 0.793 | 1.535 | 1.64 |
| 90 | 0.707 | 1.369 | 2.07 |
| 105 | 0.609 | 1.198 | 2.74 |
| 120 | 0.500 | 1.031 | 3.88 |

The last column in Table 3 shows that the increase in the energy dissipation rate using the new method is small at low impingement angles but increases significantly at higher impingement angles. The original impinging-sheet micromixing study used a fixed impingement angle ($2\beta$) equal to 30° (Demyanovich and Bourne, 1989). Since the new calculation for $\epsilon$ is only 8% higher than the old method for calculating $\epsilon$ at $2\beta = 30^o$, there is no need to revise these micromixing results.

Table 3 illustrates that if the old method were extended to calculate the energy dissipation rate at an impingement angle of 120°, $\epsilon_o$ would underestimate the energy dissipation rate calculated using the new method by almost a factor of 4. If $2\beta = 90^o$, the underestimation is a factor of 2.

The difference in the methods for calculating $\epsilon$ can have a significant impact on the calculation of the lamella size scale in which micromixing is important, particularly at high impingement angles. A concurrent micromixing study on impinging sheets used a fluid mechanical approach to analyse the results of fluorescence experiments designed to measure the size of these lamellar structures (Demyanovich, 2024). The analysis found that the lamella thickness for micromixing ($2L$) is a function of $\epsilon^{-1/2}$. Therefore, the micromixing time estimated as $L^2/D$, where $D$ is the diffusion coefficient, is proportional to $1/\epsilon$. At an impingement angle of 120°, the micromixing time would be almost 4 times higher if $\epsilon_o$ is used to estimate the energy dissipation rate instead of $\epsilon$ calculated from Eq. 17. At a 90° impingement angle, the micromixing time would be twice as high if $\epsilon_o$ is used to estimate the energy dissipation rate.

## *3.3 Derivation of the characteristics of the large-scale turbulent eddies in the impingement zone of impinging sheets*

As noted earlier, many micromixing models of systems including chemical reactions are based on concepts from fluid mechanics, and the micromixing model for impinging sheets is no exception. Besides $\epsilon$, concepts such as large-eddy size, large-eddy turnover time, fluctuating turbulent velocity, turbulent kinetic energy, turbulence intensity, large-eddy Reynolds number, and scaling of eddy sizes with large-eddy Reynolds number are of interest. However, Eq. 17 for $\epsilon$ was derived independently of turbulence theory. If the flow in the impingement zone is turbulent, then Eq. 17 can be compared with $\epsilon$ derived from turbulence energy-cascade theory and the turbulence parameters of interest can be determined.

Since the liquid sheets are thin, the question of whether turbulence is generated in the energy dissipation zone arises. The assertion is that the energy dissipation zone represents a turbulent flow. The single and mixed sheets are not considered to be turbulent flows, at least not at the velocity and thickness dimensions studied to date. From Tennekes and Lumley (1972):

1) *Turbulent flows are dissipative.* The experimental results demonstrate that the velocity is reduced from the single-sheet velocity to the mixed-sheet velocity by the time liquid reaches the end of the impingement zone.

2) *The diffusivity of turbulence causes rapid mixing and increased rates of momentum, heat and mass transfer.* Rapid micromixing resulting from the impingement of thin liquid sheets has been previously reported (Demyanovich and Bourne, 1989). If energy were not dissipated within the impingement zone, the representative size scale of the lamella in which diffusion occurs would be of order the single-sheet thickness at impingement. If this were the case, the micromixing times would be a factor of $10^2$ to $10^3$ times higher than experimentally observed.

3) *Turbulence is rotational and three dimensional.* As noted earlier, the amount of kinetic energy released by the impingement is a function of whether the collision is elastic or inelastic ($COR$ study, Demyanovich (2021b)). The demarcation between elastic and inelastic collisions depends upon whether the single-sheet velocity ($u$) is above or below a critical velocity ($u_c$). The collision is inelastic if $u < u_c$ and liquid is not ejected from the backward mixed sheet as shown in the left photograph of Fig. 11. The collision is elastic if $u > u_c$, which produces backflow as shown in the right photograph of Fig. 11.

With elastic collisions droplets are primarily ejected from rotating ligaments that are three dimensional. Other than energy dissipation which produces turbulence in the impingement zone there does not seem to be another plausible explanation for these rotating ligaments. Although such rotation cannot be visually observed for inelastic collisions, there is no reason to believe that the impingement zone is not turbulent especially since the released kinetic energy is greater for inelastic collisions than it is for elastic collisions at the same impingement angle, single-sheet thickness and single-sheet velocity ($COR$ study and Eq. 4). Although it is asserted that the flow in the impingement zone is turbulent, the Reynolds numbers for these experiments were too low to be considered fully developed turbulence, which is discussed further in Section 3.3.1.

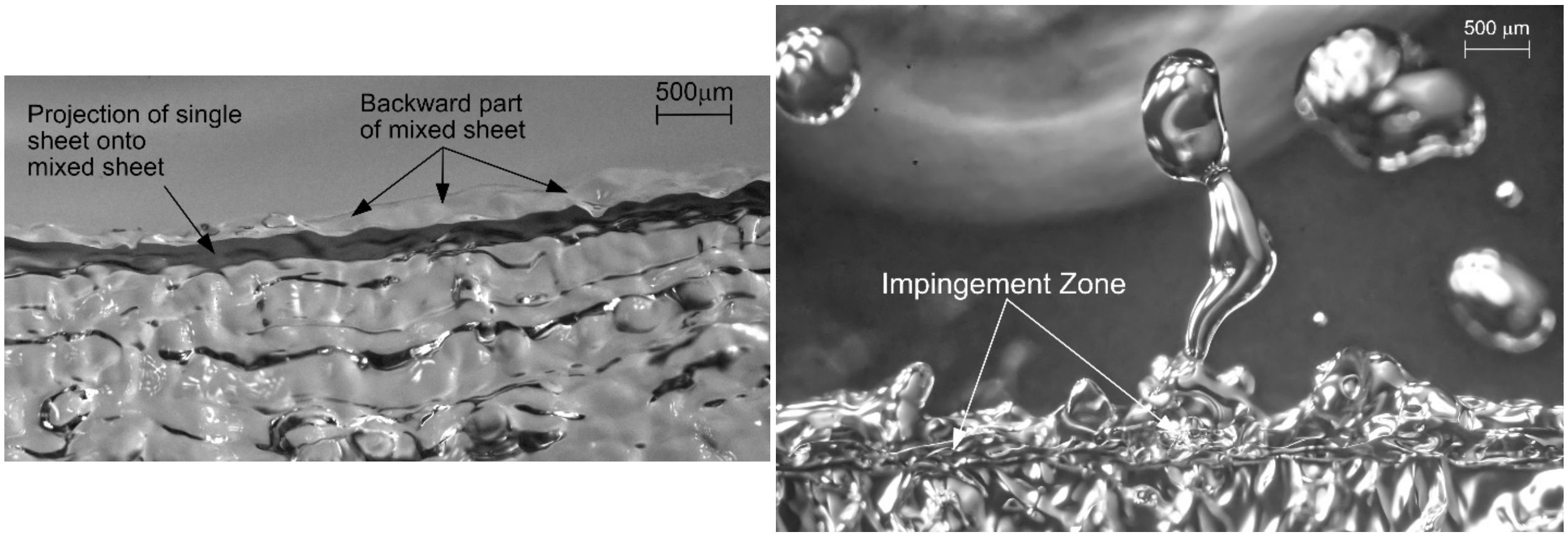

**Fig. 11.** Left photograph – 4 μs flash photograph of the front view of an inelastic collision of two equal sheets of water at an impingement angle of 90°. Right photograph - 4 μs flash photograph of the front view of an elastic collision of two equal sheets of 40% aqueous glycerin at an impingement angle of 110°. In both photographs the thickness of the single sheets at impingement is calculated as 87 μm. The right photograph shows how droplets are produced from the backward mixed sheet by the pinch-off of liquid from rotating ligaments. The left photograph is a reproduction of Fig. 2 in the $COR$ study and the right photograph is a reproduction of Fig. 12 in Demyanovich (2021b).

Eq. 17 can now be compared with energy dissipation rates derived from turbulence energy-cascade theory (Richardson, 1920), which assumes that large or integral eddies supply energy to small eddies at a rate which is inversely proportional to the large-eddy time scale. This suggests that large, energy-containing eddies pass their kinetic energy (per unit mass) to smaller eddies within their life span or turnover time (Ting, 2016).

The large-eddy turnover time ($t_\Lambda$), which varies inversely with the transfer rate of kinetic energy, is given by

$$t_{\Lambda} = \frac{\Lambda}{U_{\Lambda}} \tag{21}$$

where $\Lambda$ is a characteristic length scale of the large (energy-containing) eddies and $U_{\Lambda}$ is the characteristic velocity of the large-eddy kinetic energy. The turbulent kinetic energy is proportional to $U_{\Lambda}^2$. Thus, the energy dissipation rate is approximately given by:

$$\epsilon \sim \frac{U_{\Lambda}^2}{t_{\Lambda}} \sim \frac{U_{\Lambda}^3}{\Lambda} \tag{22}$$

An equality for Eq. 22 is achieved by including a factor, $C_{\epsilon}$,

$$\epsilon = C_{\epsilon}\frac{U_{\Lambda}^2}{t_{\Lambda}} = C_{\epsilon}\frac{U_{\Lambda}^3}{\Lambda} \tag{23}$$

where $C_{\epsilon}$ is a constant of order 1 that should be independent of Reynolds number ($Re$) (Pearson, et al., 2004). Tennekes and Lumley (1972) designate the proportionality constant as $A$, but the more recent designation of $C_{\epsilon}$ will be used in the following discussion.

The dimensionless dissipation rate coefficient, $C_{\epsilon}$, has been the subject of much research that is well summarized by Vassilicos (2015). Although $C_{\epsilon}$ should be independent of $Re$, for many turbulent flows, $C_{\epsilon}$ is a function of $Re$. Researchers have found that an asymptotic value of $C_{\epsilon}$ is reached at high $Re$. These asymptotic values, $C_{\epsilon A}$, are minimum values of $C_{\epsilon}$ when plotted versus $Re$. Pearson et al. (2004, 2002) report $C_{\epsilon A} \approx 0.5$ for decaying turbulence. McComb et al. (2015) provide a brief review of both numerical and experimental results on $C_{\epsilon}$ concluding that "the asymptotic value of $C_{\epsilon A} \cong 0.5$ is a well-established numerical result which is broadly in agreement with experimental work."

Kinetic energy from the collision of impinging sheets is dissipated in a volume equal to the impingement zone volume. Since kinetic energy is not released anywhere else in the mixed sheet, the minimum energy dissipation rate is calculated when the dissipation volume is the impingement zone volume as in Eq. 17. To compare $\epsilon$ derived from energy cascade theory with $\epsilon$ derived independently of energy cascade theory, $C_{\epsilon}$ will be set equal to the minimum, asymptotic value which is $C_{\epsilon A}$. For the moment, it will be assumed that for impinging sheets there is no dependency of $C_{\epsilon}$ on $Re$, which is theoretically expected. However, this assumption will be revisited at the end of Section 3.3.1.

With $C_{\epsilon} = C_{\epsilon A}$, setting Eq. 23 equal to Eq. 17 yields:

$$\epsilon = C_{\epsilon A}\frac{U_{\Lambda}^2}{t_{\Lambda}} = \frac{v'^2 u}{Z\Delta r} \tag{24}$$

Eq. 24 indicates that the turbulent kinetic energy ($U_\Lambda^2$) of the large eddies is $v'^2$, hence the characteristic velocity ($U_\Lambda$) of the large energy-containing eddies is equal to $v'$. The turbulence intensity ($TI$) is equal to $v'/u$.

From Eq. 24 the large-eddy turnover time is given by

$$t_\Lambda = \frac{C_{\epsilon A} Z \Delta r}{u} \tag{25}$$

In Eq. 21 the large-eddy turnover time is based on the velocity characteristic of the energy release or fluctuating velocity ($U_\Lambda = v'$) and not the velocity of the mean flow ($u$). Eq. 25 indicates that the mean flow velocity should be used for impinging sheets; however, these velocities are strongly correlated since $v'$ is a direct function of $u$ and the $COR$ via Eq. 5.

Eq. 5 allows $t_\Lambda$ to be derived using $v'$(noting that $\Delta r = s_i/sin\beta$),

$$t_\Lambda = \frac{C_{\epsilon A} Z s_i (1 - COR^2)^{1/2}}{v' sin\beta} \tag{26}$$

For an inelastic collision of impinging sheets, $COR = cos\beta$, and Eq. 26 reduces to:

$$t_\Lambda = \frac{C_{\epsilon A} Z s_i (1 - cos^2\beta)^{1/2}}{v' sin\beta} = \frac{C_{\epsilon A} Z s_i (sin^2\beta)^{1/2}}{v' sin\beta} = \frac{C_{\epsilon A} Z s_i}{v'} \tag{27}$$

For an inelastic collision $\Delta r/u = s_i/v'$ resulting in no difference in the calculation of $t_\Lambda$ using $\Delta r$ and $u$ on the one hand or $s_i$ and $v'$ on the other. For an elastic collision $s_i/v'$ is up to 36% greater than $\Delta r/u$ over the range $30^o \leq 2\beta \leq 120^o$.

The impingement zone residence time ($t_r$) for the experimental runs is listed in column 2 of Table 2. For the experimental results with $C_{\epsilon A} = 0.5$, the mean value of $t_\Lambda/t_r$ for all runs is 96% with a percent standard deviation of 1.9%. Thus, the large-eddy turnover time is approximately equal to the residence time of liquid in the impingement zone. Since the kinetic energy is dissipated within $t_r$ and $t_\Lambda \approx t_r$, the kinetic energy is dissipated approximately within $t_\Lambda$. Therefore, calculation of $t_\Lambda$ from either $\Delta r$ and $u$ in Eq. 25 or $s_i$ and $v'$ in Eq. 27 leads to a result that is in accordance with the notion from turbulence energy-cascade theory that large, energy-containing eddies lose their energy to smaller eddies within one large-eddy turnover time.

The large-eddy length scale, $\Lambda$, is typically constrained by the physical boundaries of the flow. With $C_{\epsilon A} \approx 0.5$, Eq. 25 indicates that $\Lambda$ is equal to $Z\Delta r/2$, assuming a velocity scale equal to $u$ for the large-eddy turnover time. Over the range $30^o \leq 2\beta \leq 120^o$, the ratio of $Z\Delta r/2$ to the length of the impingement zone, $\Delta b$, varies from 1.01 to 1.375 for an inelastic collision and from 0.99 to 1.04 for an elastic collision. Fig. 12a is a conceptual diagram of the large eddies if $\Lambda$ is equal to $Z\Delta r/2$, whereas Fig 12b is a conceptual diagram of the large eddies if $\Lambda = Zs_i/2$ as suggested by Eq. 27.

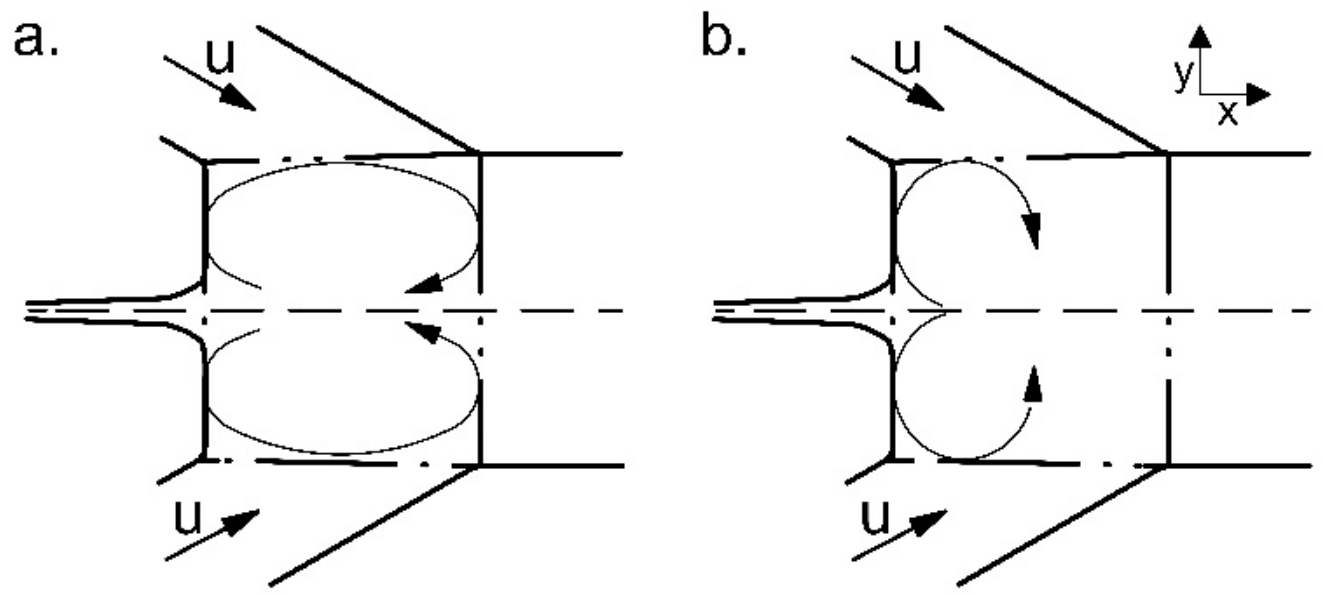

**Fig. 12.** Schematic of some possible large-eddy sizes taken from the turnover time. In Fig. a, $2\beta = 60^o, C_{\epsilon A}$ = 0.5 and $t_\Lambda = Z\Delta r/(2u)$, so $\Lambda \sim Z\Delta r/2$ and the largest eddy size is confined to about half of the impingement zone. In Fig. b, $2\beta = 60^o, C_{\epsilon A} = 0.5$ and $t_\Lambda = Zs_i/(2v')$, so $\Lambda \sim Zs_i/2$.

The analysis of the experimental results, however, only allows for the determination of the large-eddy turnover time and not the large-eddy length scale or the characteristic velocity of the turnover time. Still, the parameters of the large-eddy turnover time (either $\Delta r$ and $u$ or $s_i$ and $v'$) can be used as an initial starting point for the definition of the Reynolds number ($Re_\Lambda$) for the large, energy-containing eddies. In the concurrent study on the micromixing of impinging sheets mentioned earlier (Demyanovich, 2024), the best correlation ($R^2$ = 0.958) of the micromixing lamella size ($2L$) with $Re_\Lambda$ was obtained when $Re_\Lambda$ was defined based on the parameters of Eq. 25 ($Re_\Lambda = uC_{\epsilon A}Z\Delta r/\nu$). Defining $Re_\Lambda$ based on the parameters of Eq. 27 as $v'C_{\epsilon A}Zs_i/\nu$ resulted in a poor correlation ($R^2$ = 0.514).

### *3.3.1 Effect of viscosity on experimentally determined energy dissipation rates*

In the turbulent energy cascade, energy is supplied by large energy-containing eddies and transferred to smaller eddies, where the energy is ultimately dissipated. However, if the Reynolds number is low, some or all of the energy can be dissipated at the scale of the large eddies by viscous dissipation ($\nu v'^2/\Lambda^2$). Kinetic energy not available for transfer can lead to relatively larger lamella sizes in which micromixing is important (i.e. the energy dissipation rate would be reduced by the amount of kinetic energy lost by the large eddies to direct viscous dissipation).

The Taylor-microscale Reynolds number ($Re_\lambda = v'^2\sqrt{15/\nu\epsilon}$) can be used to determine the $Re$ of a turbulent flow. Dimotakis (2000) investigated a number of flow systems including shear layers, jets, pipes and grids and found that fully developed turbulence is achieved when $Re_\lambda \geq 100 - 140$. For the experimental data presented here, $Re_\lambda$ varied from 27 to 52 indicating that the flow in the impingement zone was not fully developed turbulence and some loss of kinetic energy to viscous dissipation might have occurred.

Since the turbulence was not fully developed, the relative magnitude of the viscous dissipation rate to the energy dissipation rate should be estimated. The experiments reported in Table 1 were conducted at kinematic viscosities of 5.3x10$^{-6}$ and 2.23x10$^{-5}$ m$^2$/s. Assuming that the largest eddies are of size $C_{\epsilon A}Z\Delta r$, the viscous dissipation rate at 5.3x10$^{-6}$ m$^2$/s is less than 1% of $\epsilon$ calculated from Eq. 17 and at 2.3x10$^{-5}$ m$^2$/s is about 3.5% of $\epsilon$.

The purpose of using higher viscosities in this study was to dampen impact waves produced as a result of the impingement. Impact waves can also be dampened to some degree by using thinner sheets at impingement. However, this can significantly reduce $\Delta b$, which, at a maximum video camera frame rate of 24,047/s, results in fewer velocity vectors to measure in the impingement zone.

Some experiments were conducted at single-sheet thicknesses of 75 µm at impingement and at kinematic viscosities of 1.0 x $10^{-6}$ $m^2/s$ and 2.0 x $10^{-6}$ $m^2/s$. The results are summarized in Table 4. Only one full impingement zone vector could be measured which meant that the measured velocity profile in the impingement zone was linear instead of nonlinear as shown in Fig. 9. For both runs, a comparison of columns 14 and 8 in Table 4 shows that the velocity of the single vector in the impingement zone was less than 2% greater than the velocity measured in the mixed sheet. For a true linear profile, however, an average of $u$ and $v_m$ results in average velocities that are 4.5% and 7.3% greater than $v_m$. Therefore, the results in Table 4 for these low-viscosity tests can be extrapolated as similar to those in Table 2 (and Fig. 9) and indicate that the results and findings were not a function of the experimental viscosity range.

Thus, at these lower viscosities, the time required for the kinetic energy to be dissipated was equal to the residence time of liquid in the impingement zone. Direct loss to viscous dissipation was insignificant since the amount of large-eddy kinetic energy lost to viscous dissipation was only 0.3% of $\epsilon$ at a kinematic viscosity equal to 1.0 x $10^{-6}$ $m^2/s$ and 0.7% of $\epsilon$ at a kinematic viscosity equal to 2.0 x $10^{-6}$ $m^2/s$.

**Table 4.** Experimental parameters for the low-viscosity experiments. Nomenclature and abbreviations are provided in Table 1. Vel. – velocity.

| Run | 2β (°) | Nom. ΔP (kPa) | $\nu$ x$10^{-6}$ ($m^2/s$) | Nbr. of feat. seq. | Nbr. of vectors meas. in IZ per seq. | Mean $u$ (m/s) | Mean $v_m$ (m/s) | % st. dev. for mean $v_m$ | Mean exp. $COR$ | cosβ | Mean $\Delta b$ (µm) | $\epsilon$ (W/kg) | Vel. of single vector in IZ (m/s) |
|---|---|---|---|---|---|---|---|---|---|---|---|---|---|
| 8 | 55 | 11.0 | 1.0 | 8 | 1 | 4.24 | 3.89 | 15.6% | 0.92 | 0.89 | 150 | 40111 | 3.96 |
| 9 | 53 | 10.0 | 2.0 | 13 | 1 | 3.69 | 3.22 | 5.4% | 0.87 | 0.89 | 142 | 40767 | 3.23 |

Finally, a few comments about the dependence of $C_\epsilon$ on $Re$. Since the change in velocity from $u$ to $v_m$ occurs within $t_r$, the kinetic energy released by the collision is dissipated within $t_r$. Since $t_\Lambda \approx t_r$, the results indicate that large scale eddies transferred their energy to smaller eddies within the impingement zone and died out within about one turnover time of the large eddies. According to Vassilicos (2015), the dependence of $C_\epsilon$ on $Re$ is a result of a significant non-equilibrium region existing in various turbulent flows causing $C_\epsilon$ to be a function of the ratio of $Re$ at the inlet of the flow and the local turbulence $Re$. This disparity in large-scale eddy Reynolds numbers does not exist within the impingement zone of impinging sheets. There is only one location for the large-scale eddies (and $Re_\Lambda$) and that location is in the impingement zone. Therefore, in accordance with theory, no dependence upon $Re$ is expected.

### *3.4 Kinetic energy profile in the energy dissipation zone*

The turbulence kinetic energy (TKE) created in the impingement zone of impinging sheets decays. Determination of how the TKE decays is of interest and should not be assumed to be similar to other systems. For grid-produced turbulence, for example, the turbulent kinetic energy follows a power-law decay with time of the form, $TKE \sim t^n$. Investigators have found that $n$ ranges from $-1.0$ to $-10/7$ (Ting, 2016).

The TKE is equal to the large-eddy turbulent kinetic energy per unit mass, per unit time ($\hat{v}'^2$). As the large-eddy turbulent kinetic energy is dissipated, the flow velocity decreases from $u$ to $v_m$. $\hat{v}'^2$ at any point in the impingement zone can be calculated, in a similar manner to substituting Eq. 2 into Eq. 5 ($v'^2 = u^2 - v_m^2$), as the difference between the kinetic energy associated with the flow velocity in the impingement zone and the kinetic energy associated with the mixed-sheet velocity, $v_m$,

$$\hat{v}'^2 = (\hat{u}^2 - v_m^2) \tag{28}$$

where $\hat{u}$ is the flow velocity in the radial direction in the impingement zone ($v_m \leq \hat{u} \leq u$). At the beginning of the impingement zone $\hat{u}$ was taken as the experimental single-sheet velocity, $u$. At the end of the impingement zone, $\hat{u}$ was taken to be equal to the experimental values of the mixed-sheet velocity, $v_m$. The values of $\hat{u}$ within the impingement zone were calculated from the derivative of the hyperbolic decline equation (Eq. 9). The hyperbolic decline equation was chosen for all runs because it yielded the highest regression coefficients of subsequent fits of an exponential equation (Eq. 29) to the energy profile. Fig. 13 plots $\hat{v}'^2$ in the impingement zone as a function of time for Runs 6 and 5.

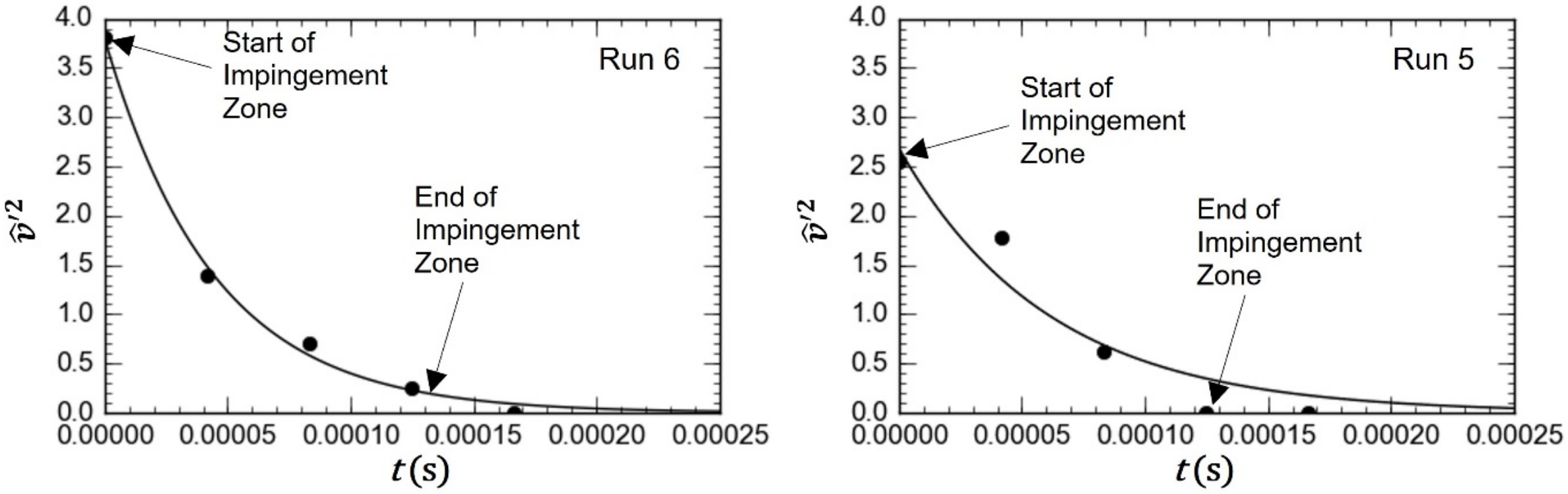

**Fig. 13.** Change in large-eddy kinetic energy within the impingement zone for Runs 6 and 5. Run 6 illustrates the fit of a typical run to an exponential equation ($R^2$ = 0.997). Run 5 yields the worst fit to an exponential equation ($R^2$ = 0.930). ● - kinetic energy calculated from Eq. 28 with $\hat{u}$ within the impingement zone determined from Eq. 9. —— exponential decay (Eq. 29).

For all experimental runs, the kinetic energy profiles were well fit by an exponential decay equation of the form

$$\hat{v}'^2 = a\exp(bt) \tag{29}$$

All regression coefficients ($R^2$) were ≥ 0.98 except for run 5 where $R^2$ = 0.93.

The constant $a$ in Eq. 29 is equal to $v'^2$ with $v'$ calculated from Eq. 5. As such, it is anticipated that the constant $b$ might correlate with the rate of transfer of kinetic energy (T). From Eq. 23, T is equal to (with $C_\epsilon = C_{\epsilon A}$)

$$\mathrm{T} = \frac{C_{\epsilon A}}{t_\Lambda} \tag{30}$$

The constant $b$ in Eq. 29 is plotted against T in Fig. 14. A linear fit to the experimental data results in

$$b = -5.10\mathrm{T} \tag{31}$$

with $R^2$ = 0.84. For the linear fit of Eq. 31, the y-intercept was forced to 0; however, without forcing (y-intercept = 20.5), the accuracy of the fit is insignificantly better ($R^2$ = 0.8392195 vs $R^2$ = 0.8392190).

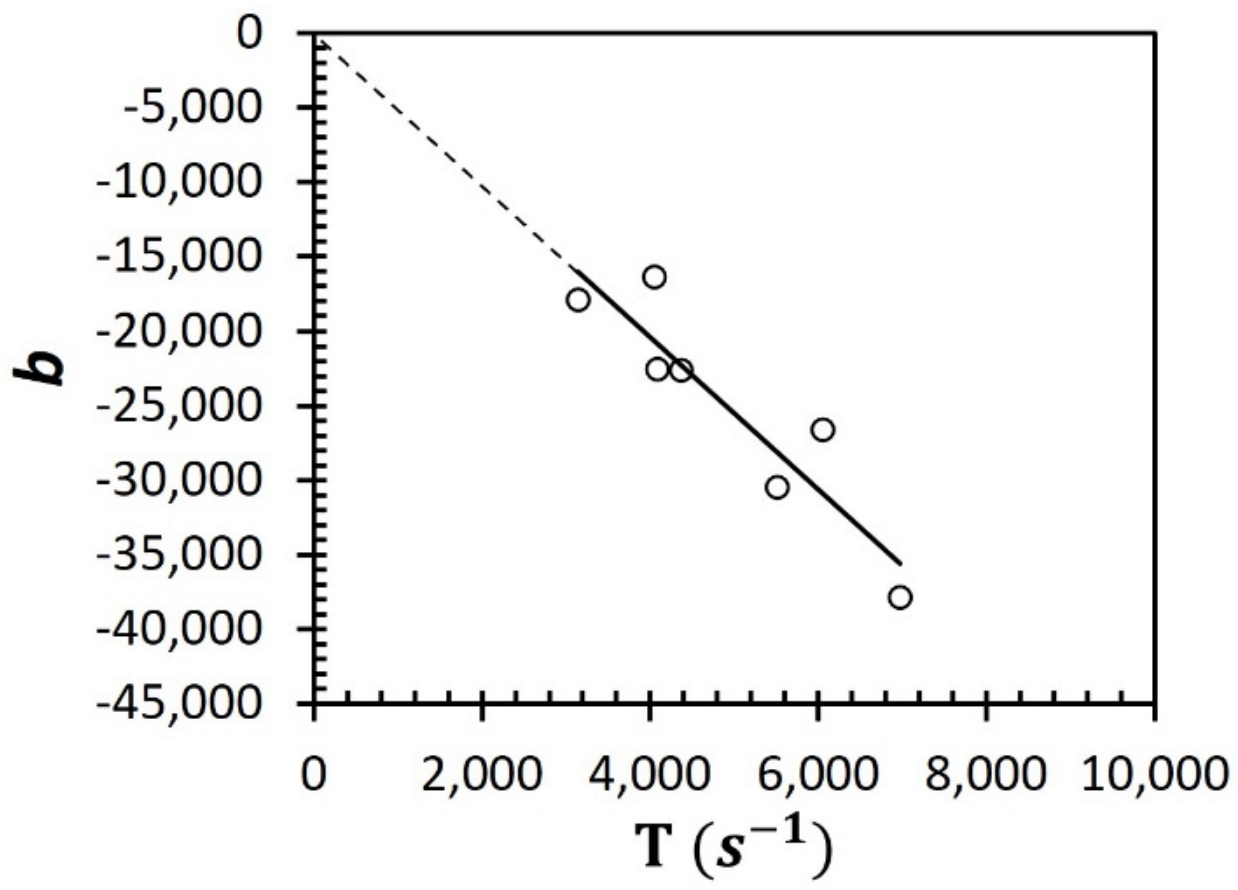

**Fig. 14.** Plot of the constant b in Eq. 29 as a function of the transfer rate of kinetic energy ($\mathrm{T} = C_{\epsilon A}/t_\Lambda$) undergoing dissipation. —— Linear fit.

Substituting Eq. 5, Eq. 31 and Eq. 30 into Eq. 29 yields

$$\hat{v}'^2 = v'^2\exp\left(-5.10\frac{C_{\epsilon A}}{t_\Lambda}t\right) = u^2(1 - COR^2)\exp\left(-\frac{2.55}{t_\Lambda}t\right) \tag{32}$$

where $C_{\epsilon A}$ is taken as 0.5 in the numerator of the quotient.

Eq. 32 can also be written as

$$\frac{\hat{v}'^2}{v'^2} = \exp\left(-\frac{2.55}{t_\Lambda}t\right) = \exp\left(-\frac{t}{\tau}\right) \tag{33}$$

where $1/\tau$ is the decay rate constant and $\tau$ is the mean lifetime equal to

$$\tau = \frac{t_\Lambda}{2.55} \tag{34}$$

$\tau$ is the time at which the large-eddy kinetic energy is reduced to $1/e$ or 36.8% of the initial value ($v'^2$). The half-life of the large-eddy kinetic energy is equal to $\tau$ ln2 or ~ $t_\Lambda/4$. For the experimental data, Eq. 33 captured 93% $\pm 0.4\%$ of the dissipation of kinetic energy that occurred by the time liquid reached the end of the impingement zone.

## 4. Summary and conclusions

The energy dissipation rate ($\epsilon$) is an important parameter for the analysis of the micromixing possible by impinging thin liquid sheets. The kinetic energy released by the impingement was previously investigated and found to be a function of the coefficient of restitution ($COR$) of the collision (Demyanovich, 2021a). The purpose of this work was to study the volume within which the released energy is dissipated.

The volume of the energy dissipation zone was investigated by measuring the velocity of holes generated in the single sheets as the holes passed through the impingement zone and into the mixed sheet. The experimental results showed that the single-sheet velocity ($u$), which is constant, was reduced to the mixed-sheet velocity ($v_m$), which is also constant, by the time liquid exited the impingement zone. The time required for the velocity to change from $u$ to $v_m$ was equal to the residence time of liquid in the impingement zone. Thus, the volume of the energy dissipation zone is equal to the volume of the impingement zone, which can be determined from geometric considerations and the $COR$ of the collision.

The formula for the calculation of $\epsilon$ was revised and compared with an earlier derivation ($\epsilon_o$). The new method results in higher calculated values of the energy dissipation rate, with $\epsilon/\epsilon_o$ increasing with increasing impingement angle. At an impingement angle of 90°, $\epsilon$ is twice as high as $\epsilon_o$ and at 120° it is almost four times higher.

Since the micromixing analysis of impinging sheets employs a fluid mechanical approach, comparisons of the derived energy dissipation rate (Eq. 17) were made with energy dissipation rates determined from turbulence energy-cascade theory (Eq. 23). The comparisons indicate that the characteristic velocity of the large-eddy kinetic energy ($U_\Lambda$) is equal to the velocity ($v'$) associated with the kinetic energy released as a result of the collision. The turbulent kinetic energy is, therefore, equal to $v'^2$ and the turbulence intensity is equal to $v'/u$. An equation for the large-eddy turnover time ($t_\Lambda$) was derived but the characteristic length scale of the large eddies ($\Lambda$) could not be definitively determined since there were two equivalent ways of deriving $t_\Lambda$.

The large eddy turnover time was found to be approximately equal to the residence time of liquid in the impingement zone ($t_r$). Since the kinetic energy was dissipated within $t_r$, and $t_\Lambda \approx t_r$ ,the large eddies lost their energy within $t_\Lambda$. This finding is in accordance with turbulence energy cascade theory which assumes that large, energy-containing eddies lose their energy to smaller eddies within $t_\Lambda$.

The profile for the loss of large-eddy kinetic energy in the energy dissipation/impingement zone was found to follow an exponential decay with a decay rate constant equal to $\frac{2.55}{t_\Lambda}$. For the experimental data, the exponential decay equation captured 93% of the loss of large-eddy kinetic energy that occurred by the time liquid had reached the end of the impingement zone.

## Declaration of competing interest

The author declares that he has no known competing financial interests or personal relationships that could have appeared to influence the work reported in this paper.

## Data availability

The data that support the findings of this study are available from the author upon reasonable request. These data do not include videos.

## Acknowledgments

The author would like to thank Professor David Ting, Mechanical, Automotive & Materials Engineering, University of Windsor for helpful discussions on turbulence theory.

## Appendix A. Derivation of Eq. 15

In the impingement zone $\dot{m} = \rho\varphi R s\hat{u}$ and since $\Delta b \ll R_i$, there is no significant change in $R$. Since $\dot{m}, \rho, \varphi$ and $R$ are constant, $s\hat{u}$ is constant. As $\hat{u}$ decreases, the thickness of the combined sheets increases in the impingement zone. The lines with a measurement of $\Delta c$ in Fig. 10b are depicted as straight lines, but, as illustrated in Fig. A.1a, are actually the inverse curve of the velocity profile ($\hat{u}$) in the impingement zone, which is similar to the profile shown in Fig. 9. However, the simplification shown in Fig. 10b and Fig. A.1b is estimated to have little effect on the calculated volume of the impingement zone.

The cross-sectional area of the impingement zone, $A_{IZ}$, can be calculated as the area of a trapezoid depicted by the gray shaded area in Fig. A.1b:

$$A_{IZ} = \frac{(s_m + s_f)}{2} \Delta b \tag{A.1}$$

Since the energy dissipation volume is equal to the impingement zone volume, the reduction in velocity to $v_m$ occurs at the end of the impingement zone (after a distance equal to $\Delta b$). If the backward mixed sheet is short relative to $R_i$, the thickness, $s_f$, is given by ($COR$ study),

$$s_f = s_i \left( \frac{COR + cos\beta}{COR^2} \right) = Bs_i \tag{A.2}$$

where $B$ is equal to $(COR + cos\beta)/COR^2$.

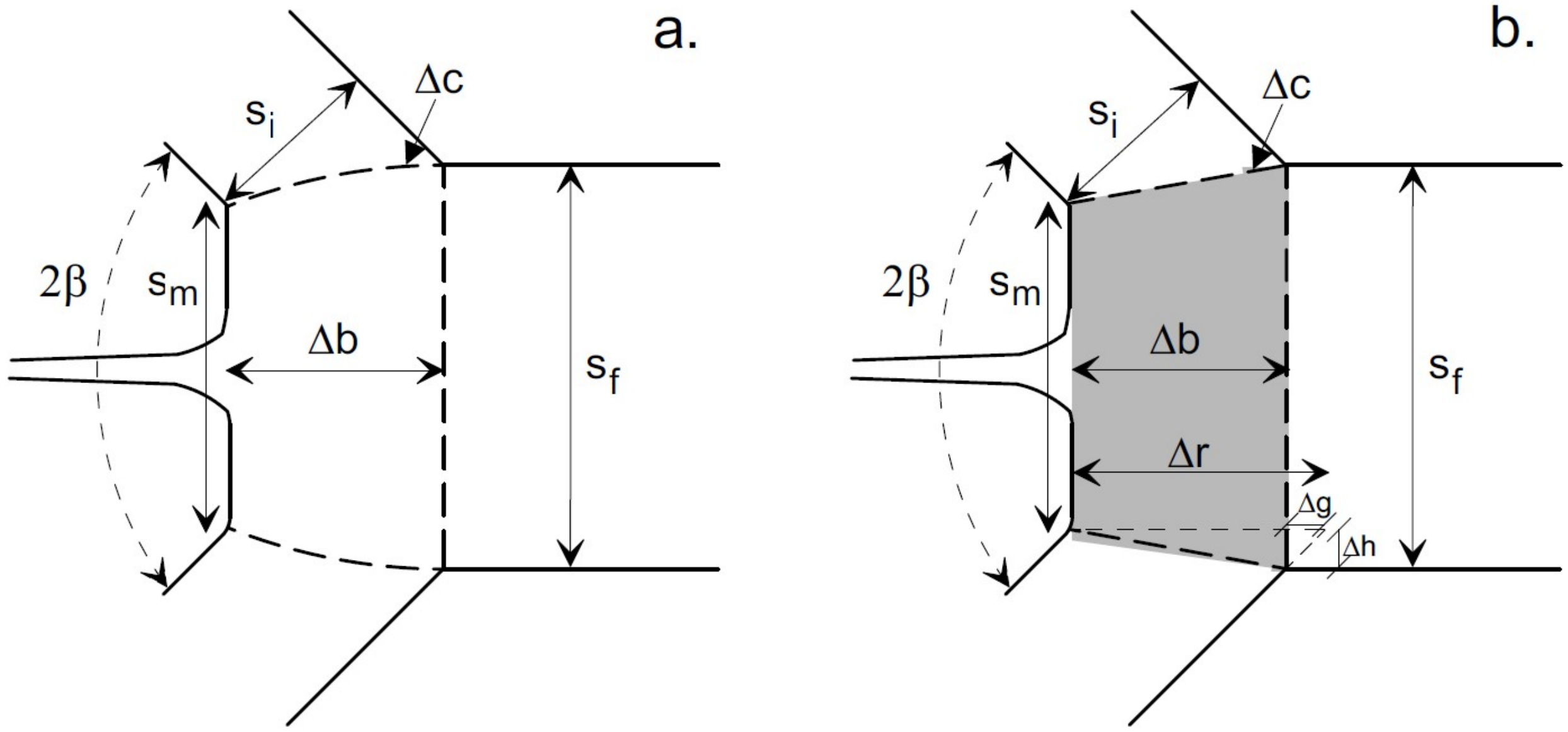

**Fig. A.1.** Schematic illustration of the impingement zone used to calculate $A_{IZ}$.

At the beginning of the impingement zone the velocity ($\hat{u}$) is equal to $u$, and the $COR$ at this point can be thought of as equal to 1. The thickness, $s_m$, can then be estimated from Eq. A.2 as

$$s_m = s_i(1 + cos\beta) \tag{A.3}$$

Since the flowrate is constant but $\hat{u}$ decreases from $u$ to $v_m$, the forward mixed sheet increases in thickness within the energy dissipation zone yielding a value of $s_f$ that is greater than $s_m$.

Substituting in for $s_f$ and $s_m$ in Eq. A.1 yields

$$A_{IZ} = \frac{(s_i + s_i cos\beta + Bs_i)}{2} \Delta b = s_i \Delta b \frac{(1 + cos\beta + B)}{2} \tag{A.4}$$

A convenient length scale to use as a reference for the length of the impingement zone ($\Delta b$) is the projection of the single sheet defined using $sin\beta$ only, which is

$$\Delta r = \frac{s_i}{sin\beta} \quad \text{(A.5)}$$

As a function of $\Delta r$, $\Delta b$ can be determined from the geometry of Fig. A.1b as follows:

$$\Delta b = \Delta r - \Delta g = \Delta r - \frac{\Delta h}{tan\beta} \quad \text{(A.6)}$$

$$\Delta h = \frac{s_f - s_m}{2} \quad \text{(A.7)}$$

therefore,

$$\Delta b = \Delta r - \frac{s_f}{2tan\beta} + \frac{s_m}{2tan\beta} \quad \text{(A.8)}$$

Plugging in Eq. A.2 for $s_f$ and Eq. A.3 for $s_m$ yields

$$\Delta b = \Delta r - \frac{Bs_i cos\beta}{2sin\beta} + \frac{s_i cos\beta}{2sin\beta} + \frac{s_i cos^2\beta}{2sin\beta} \quad \text{(A.9)}$$

Using Eq. A.5, Eq. A.9 simplifies to

$$\Delta b = \Delta r\left(1 - \frac{Bcos\beta}{2} + \frac{cos\beta}{2} + \frac{cos^2\beta}{2}\right) = F\Delta r \quad \text{(A.10)}$$

Over the range $30^o \leq 2\beta \leq 120^o$, $F$ varies from 0.967 to 0.707 for an elastic collision and 0.949 to 0.375 for an inelastic collision.

Eq. A.4 now becomes

$$A_{IZ} = s_i F\Delta r\left(\frac{1 + cos\beta + B}{2}\right) \quad \text{(A.11)}$$

or, substituting in for $F$ and multiplying out results in

$$A_{IZ} = s_i\Delta r\left(\frac{2 + 3cos\beta + 2cos^2\beta + cos^3\beta + 2B - B^2cos\beta}{4}\right) \quad \text{(A.12)}$$

The final simplified equation for the cross-sectional area of the impingement zone is

$$A_{IZ} = Zs_i\Delta r \quad \text{(A.13)}$$

with $Z$ equal to

$$Z = \frac{2 + 3cos\beta + 2cos^2\beta + cos^3\beta + 2B - B^2cos\beta}{4} \quad \text{(A.14)}$$

and

$$B = \frac{COR + cos\beta}{COR^2} \quad \text{(A.15)}$$

## Appendix B. Propagation of uncertainty in the calculation of the energy dissipation rate

The uncertainty in the calculation of the energy dissipation rate ($\epsilon$) results from the effects on the calculation of $\epsilon$ by the uncertainty in the variables. To estimate the uncertainty in $\epsilon$, Eq. 17 is rewritten in terms of the variables that were measured,

$$\epsilon = \frac{P}{\rho V} = \frac{\rho\varphi R_i s_i u^3(1 - COR^2)}{\rho\varphi R_i A_{IZ}} = \frac{s_i u^3(1 - COR^2)}{s_i F\Delta r\left(\frac{1 + cos\beta + B}{2}\right)} \quad \text{(B.1)}$$

where Eq. A.11 is substituted for $A_{IZ}$. Substituting for $F\Delta r$ using Eq. A.10 results in

$$\epsilon = \frac{2u^3(1 - COR^2)}{\Delta b(1 + cos\beta + B)} \quad \text{(B.2)}$$

The propagation of error in the calculation of $\epsilon$ is then due to the uncertainty in the variables, $u$, $COR$, $\Delta b$, $\beta$ and $B$. The standard deviations of the measurement of the single-sheet velocity ($u$) and the length of the impingement zone ($\Delta b$) are given in Table 1. The error in the measurement of the angle of impingement ($2\beta$) is typically 1°. The uncertainty in the calculation of the $COR$ in Eq. 2 was based on the standard deviations of $u$ and $v_m$ given in Table 1. The uncertainty in the calculation of $B$ in Eq. A.15 was based on the propagated uncertainty for the calculation of the $COR$ and the estimated uncertainty in the measurement of $\beta$. The propagated uncertainty in the calculation of the $COR$ and $B$ are given by

$$\sigma_{COR} = \sqrt{\frac{v_m^2 \sigma_u^2}{u^4} + \frac{\sigma_{v_m}^2}{u^2}} \tag{B.3}$$

$$\sigma_B = \sqrt{\sigma_{COR}^2 \left( \frac{1}{COR^2} - \frac{2(COR + cos\beta)}{COR^3} \right)^2 + \frac{\sigma_{cos\beta}^2}{COR^4}} \tag{B.4}$$

where $\sigma_u$ and $\sigma_{v_m}$ are the uncertainties in $u$ and $v_m$, and $\sigma_{COR}$ and $\sigma_B$ are the propagated uncertainties in the calculations of the $COR$ and $B$, respectively. The uncertainty in the calculation of $cos\beta$ is given by,

$$\sigma_{cos\beta} = \sigma_\beta \sqrt{sin^2\beta} \tag{B.5}$$

where $\sigma_\beta$ is the uncertainty in the measurement of $\beta$. The uncertainty in the calculation of the energy dissipation rate, $\sigma_\epsilon$, can then be calculated from the following equation,

$$\sigma_\epsilon = \sqrt{\frac{16COR^2\sigma_{COR}^2 u^6}{\Delta b^2(B + \cos(\beta) + 1)^2} + \frac{4\sigma_B^2 u^6(1 - COR^2)^2}{\Delta b^2(B + \cos(\beta) + 1)^4} + \frac{4\sigma_{cos\beta}^2 u^6(1 - COR^2)^2}{\Delta b^2(B + \cos(\beta) + 1)^4} + \frac{36\sigma_u^2 u^4(1 - COR^2)^2}{\Delta b^2(B + \cos(\beta) + 1)^2} + \frac{4\sigma_{\Delta b}^2 u^6(1 - COR^2)^2}{\Delta b^4(B + \cos(\beta) + 1)^2}} \tag{B.6}$$

Table B.1 lists the calculated values of $\epsilon$ for each run along with the estimated uncertainty in the calculation of $\epsilon$.

**Table B.1.** Calculation of the energy dissipation rate, $\epsilon$, from Eq. B.2 and the propagated uncertainty, $\sigma_\epsilon$, from Eq. B.6.

| Run | $u$ (m/s) | $\sigma_u$ (m/s) | $COR$ | $\sigma_{COR}$ | $\Delta b$ (µm) | $\sigma_{\Delta b}$ (µm) | $cos\beta$ | $\sigma_{cos\beta}$ | $B$ | $\sigma_B$ | $\epsilon$ (m²/s³) | $\sigma_\epsilon$ (m²/s³) | $\frac{\sigma_\epsilon}{\epsilon}$ |
|---|---|---|---|---|---|---|---|---|---|---|---|---|---|
| 1 | 3.56 | 0.034 | 0.91 | 0.057 | 312 | 31.2 | 0.92 | 0.0068 | 2.23 | 0.21 | $1.25x10^4$ | $7.25x10^3$ | 57.8% |
| 2 | 4.57 | 0.055 | 0.89 | 0.034 | 361 | 26.7 | 0.92 | 0.0068 | 2.27 | 0.13 | $2.54x10^4$ | $7.98x10^3$ | 31.5% |
| 3 | 3.37 | 0.047 | 0.73 | 0.062 | 333 | 22.0 | 0.79 | 0.011 | 2.84 | 0.37 | $2.30x10^4$ | $5.21x10^3$ | 22.6% |
| 4 | 2.75 | 0.068 | 0.74 | 0.089 | 379 | 34.1 | 0.79 | 0.011 | 2.82 | 0.52 | $1.09x10^4$ | $3.58x10^3$ | 32.9% |
| 5 | 3.08 | 0.071 | 0.86 | 0.064 | 367 | 37.1 | 0.83 | 0.0096 | 2.31 | 0.26 | $1.03x10^4$ | $4.44x10^3$ | 42.9% |
| 6 | 3.26 | 0.023 | 0.81 | 0.049 | 368 | 25.0 | 0.83 | 0.0096 | 2.50 | 0.23 | $1.49x10^4$ | $3.69x10^3$ | 24.9% |
| 7 | 3.95 | 0.042 | 0.79 | 0.044 | 255 | 15.6 | 0.83 | 0.0096 | 2.60 | 0.22 | $4.09x10^4$ | $8.30x10^3$ | 20.3% |
| | | | | | | | | | | | | **Mean:** | **33.3%** |

The mean relative uncertainty for all runs is 33%. Thus, the energy dissipation rate calculated from Eq. 17 or Eq. B.2 has an uncertainty of about one-third. A portion of this uncertainty is due to the variation in $\Delta b$ at different azimuthal locations within the impingement zone. Uncertainty in the value of the $COR$ has a large impact, since it also creates uncertainty in the value of $B$ (see Eq. B.2).

As noted earlier, the micromixing time can be approximately estimated as $L^2/D$, where $L$ is the half lamella size in which micromixing (diffusion) is important and $D$ is the diffusion coefficient. For low-viscosity liquids, $D$ is of the order of $10^{-9}$ $m^2/s$. For impinging sheets, energy dissipation rates of $10^5$ W/kg are easily achieved. At this $\epsilon$ and at single sheet flow rates of 0.75 to 1.2 L/min, the calculated value of $L$ is of order 6 to 7 μm leading to an estimated micromixing time of < 0.05 s. If the energy dissipation rate is one-third higher or one-third lower, this micromixing time varies by about -20% or +20%, respectively.

## References

Abiev, R.S., Sirotkin, A.A., 2020. Influence of hydrodynamic conditions on micromixing in microreactors with free impinging jets. Fluids 5. https://doi.org/10.3390/fluids5040179

Baldyga, J., Bourne, J.R., 1984a. A Fluid Mechanical Approach to Turbulent Mixing and Chemical Reaction Part III Computational and Experimental Results for the New Micromixing Model. Chem Eng Commun 28, 259–281. https://doi.org/10.1080/00986448408940137

Baldyga, J., Bourne, J.R., 1984b. A Fluid Mechanical Approach to Turbulent Mixing and Chemical Reaction Part II Micromixing in the Light of Turbulence Theory. Chem Eng Commun 28, 243–258. https://doi.org/10.1080/00986448408940136

Baldyga, J., Bourne, J.R., 1984c. A Fluid Mechanical Approach to Turbulent Mixing and Chemical Reaction Part I Inadequacies of Available Methods. Chem Eng Commun 28, 231–241. https://doi.org/10.1080/00986448408940135

Bourne, J.R., 2003. Mixing and the selectivity of chemical reactions. Org Process Res Dev 7, 471–508. https://doi.org/10.1021/OP020074Q

Brown, R.H., Schneider, S.C., Mulligan, M.G., 1992. Analysis of algorithms for velocity estimation from discrete position versus time data. IEEE Transactions on Industrial Electronics 39, 11–19. https://doi.org/10.1109/41.121906

Clanet, C., Villermaux, E., 2002. Life of a smooth liquid sheet. J Fluid Mech 462, 307–340. https://doi.org/10.1017/S0022112002008339

Demyanovich, R., 2024. Experimental Study and Turbulence Dissipative Scale Modelling of the Rapid Micromixing of Impinging, Paper-Thin Sheets of Liquids. ChemRxiv [Preprint]. January 30, 2024. Available from: https://chemrxiv.org/engage/chemrxiv/article-details/65b9873ee9ebbb4db95077df

Demyanovich, R.J., 2021a **($\boldsymbol{COR}$ study)**. High energy dissipation rates from the impingement of free paper-thin sheets of liquids: A study of the coefficient of restitution of the collision. Chemical Engineering Science: X 12, 100113. https://doi.org/10.1016/J.CESX.2021.100113

Demyanovich, R.J., 2021b. On the impingement of free, thin sheets of liquids - A photographic study of the impingement zone. AIP Adv 11. https://doi.org/10.1063/5.0040336

Demyanovich, R.J., 1991a. Production of commercial dyes via impingement-sheet mixing. Part I. Testing of a device suitable for industrial application. Chemical Engineering and Processing 29, 173–177. https://doi.org/10.1016/0255-2701(91)85017-I

Demyanovich, R.J., 1991b. Production of commercial dyes via impingement-sheet mixing Part II. Results of laboratory experiments. Chemical Engineering and Processing 29, 179–183. https://doi.org/10.1016/0255-2701(91)85018-J

Demyanovich, R.J., 1991c. Absorption of carbon dioxide by impinging, thin liquid sheets. Chem Eng Commun 103, 151–166. https://doi.org/10.1080/00986449108910868

Demyanovich, R.J., 1988. Liquid mixing employing expanding, thinning liquid sheets. U.S. patent 4,735,359. U.S. Pat. 4,735,359.

Demyanovich, R.J., Bourne, J.R., 1989. Rapid Micromixing by the Impingement of Thin Liquid Sheets. 2. Mixing Study. Ind Eng Chem Res 28, 830–839. https://doi.org/10.1021/ie00090a027

Dimotakis, P.E., 2000. The mixing transition in turbulent flows. J Fluid Mech 409, 69–98. https://doi.org/10.1017/S0022112099007946

Dombrowski, N., Hasson, D., Ward, D.E., 1960. Some aspects of liquid flow through fan spray nozzles. Chem Eng Sci 12, 35–50. https://doi.org/10.1016/0009-2509(60)90004-X

Dombrowski, N., Hooper, P.C., 1962. The performance characteristics of an impinging jet atomizer in atmospheres of high ambient density. Fuel 41, 323–334.

Gomes, N.M.O., Fonte, C.P., Sousa, C.C. e., Mateus, A.J., Bártolo, P.J., Dias, M.M., Lopes, J.C.B., Santos, R.J., 2016. Real time control of mixing in Reaction Injection Moulding. Chemical Engineering Research and Design 105, 31–43. https://doi.org/10.1016/j.cherd.2015.10.042

Halls, B.R., Meyer, T.R., Kastengren, A.L., 2015. Liquid mixing in doublet impinging jet injectors using x-ray fluorescence, in: ILASS Americas, 27th Annual Conference on Liquid Atomization and Spray Systems. pp. 1–11.

Lin, S.P., 2003. Breakup of Liquid Sheets and Jets, Breakup of Liquid Sheets and Jets. Cambridge University Press. https://doi.org/10.1017/cbo9780511547096

Mahajan, A.J., Kirwan, D.J., 1996. Micromixing Effects in a Two-Impinging-Jets Precipitator. AIChE Journal 42, 1801–1814. https://doi.org/10.1002/aic.690420702

McComb, W.D., Berera, A., Yoffe, S.R., Linkmann, M.F., 2015. Energy transfer and dissipation in forced isotropic turbulence. Phys Rev E Stat Nonlin Soft Matter Phys 91. https://doi.org/10.1103/PhysRevE.91.043013

Ottino, J.M., Ranz, W.E., Macosko, C.W., 1979. A lamellar model for analysis of liquid-liquid mixing. Chem Eng Sci 34, 877–890. https://doi.org/10.1016/0009-2509(79)85145-3

Pearson, B.R., Krogstad, P.Å., Van De Water, W., 2002. Measurements of the turbulent energy dissipation rate. Physics of Fluids 14, 1288–1290. https://doi.org/10.1063/1.1445422

Pearson, B.R., Yousef, T.A., Haugen, N.E.L., Brandenburg, A., Krogstad, P.Å., 2004. Delayed correlation between turbulent energy injection and dissipation. Phys Rev E Stat Phys Plasmas Fluids Relat Interdiscip Topics 70, 6. https://doi.org/10.1103/PhysRevE.70.056301

Richardson, M.L.F., 1920. The supply of energy from and to atmospheric eddies. Proceedings of the Royal Society of London. Series A, Containing Papers of a Mathematical and Physical Character 97, 354–373. https://doi.org/10.1098/RSPA.1920.0039

Schneider, C.A., Rasband, W.S., Eliceiri, K.W., 2012. NIH Image to ImageJ: 25 years of image analysis. Nat Methods 9, 671–675. https://doi.org/10.1038/nmeth.2089

Tennekes, H., Lumley, J.L., 1972. A First Course in Turbulence, A First Course in Turbulence. MIT Press. https://doi.org/10.7551/mitpress/3014.001.0001

Ting, D.S.-K., 2016. Basics of Engineering Turbulence. Academic Press.

Tucker, C.L., Suh, N.P., 1980. Mixing for reaction injection molding. I. Impingement mixing of liquids. Polym Eng Sci 20, 875–886. https://doi.org/10.1002/PEN.760201307

Vassilicos, J.C., 2015. Dissipation in turbulent flows. Annu Rev Fluid Mech 47, 95–114. https://doi.org/10.1146/annurev-fluid-010814-014637

Villermaux, E., Pistre, V., Lhuissier, H., 2013. The viscous Savart sheet. J Fluid Mech 730, 607–625. https://doi.org/10.1017/jfm.2013.354
Villermaux, J., Blavier, L., 1984. Free radical polymerization engineering-I. A new method for modeling free radical homogeneous polymerization reactions. Chem Eng Sci 39, 87–99. https://doi.org/10.1016/0009-2509(84)80133-5